\documentclass[%
reprint,
superscriptaddress,
 amsmath,amssymb,
 aps,
 prc,
 showkeys,
 preprintnumbers,
floatfix,
]{revtex4-2}

\usepackage{graphicx}
\usepackage{dcolumn}
\usepackage{bm}
\usepackage{amsmath}
\usepackage{comment}
\usepackage{multirow}
\usepackage[colorlinks=true,linkcolor=blue,citecolor=blue,urlcolor=blue]{hyperref}
\usepackage{cleveref}
\crefname{equation}{Eq.}{Eqs.}
\crefname{figure}{Fig.}{Figs.}
\usepackage{url}
\usepackage{float}

\def\Qt2{{Q}^2}
\def\MM2{M_X^2}

\begin{document}

\title{Measurement of the near-threshold J$/\psi$ photoproduction cross section with the CLAS12 experiment}

\newcommand*{\ANL}{Argonne National Laboratory, Argonne, Illinois 60439, USA}
\newcommand*{\ANLindex}{1}
\affiliation{\ANL}
\newcommand*{\ASU}{Arizona State University, Tempe, Arizona 85287, USA}
\newcommand*{\ASUindex}{2}
\affiliation{\ASU}
\newcommand*{\CSTATE}{California State University, Dominguez Hills, Carson, California 90747, USA}
\newcommand*{\CSTATEindex}{3}
\affiliation{\CSTATE}
\newcommand*{\CANISIUS}{Canisius College, Buffalo, New York 14208, USA}
\newcommand*{\CANISIUSindex}{4}
\affiliation{\CANISIUS}
\newcommand*{\SACLAY}{IRFU, CEA, Universit\'{e} Paris-Saclay, F-91191 Gif-sur-Yvette, France}
\newcommand*{\SACLAYindex}{5}
\affiliation{\SACLAY}
\newcommand*{\CNU}{Christopher Newport University, Newport News, Virginia 23606, USA}
\newcommand*{\CNUindex}{6}
\affiliation{\CNU}
\newcommand*{\UCONN}{University of Connecticut, Storrs, Connecticut 06269, USA}
\newcommand*{\UCONNindex}{7}
\affiliation{\UCONN}
\newcommand*{\DUKE}{Duke University, Durham, North Carolina 27708, USA}
\newcommand*{\DUKEindex}{8}
\affiliation{\DUKE}
\newcommand*{\DUQUESNE}{Duquesne University, Pittsburgh, Pennsylvania 15282, USA}
\newcommand*{\DUQUESNEindex}{9}
\affiliation{\DUQUESNE}
\newcommand*{\FU}{Fairfield University, Fairfield, Connecticut 06824, USA}
\newcommand*{\FUindex}{10}
\affiliation{\FU}
\newcommand*{\FERRARAU}{Universit\`{a} di Ferrara , 44121 Ferrara, Italy}
\newcommand*{\FERRARAUindex}{11}
\affiliation{\FERRARAU}
\newcommand*{\FIU}{Florida International University, Miami, Florida 33199, USA}
\newcommand*{\FIUindex}{12}
\affiliation{\FIU}
\newcommand*{\Genova}{Dipartimento di Fisica dell’Universit\`{a}, 16146 Genova, Italy}
\newcommand*{\Genovaindex}{13}
\affiliation{\Genova}
\newcommand*{\GWUI}{The George Washington University, Washington, D.C. 20052, USA}
\newcommand*{\GWUIindex}{14}
\affiliation{\GWUI}
\newcommand*{\GSIFFN}{GSI Helmholtzzentrum fur Schwerionenforschung GmbH, D-64291 Darmstadt, Germany}
\newcommand*{\GSIFFNindex}{15}
\affiliation{\GSIFFN}
\newcommand*{\ORSAY}{Universit\'{e} Paris-Saclay, CNRS/IN2P3, IJCLab, 91405 Orsay, France}
\newcommand*{\ORSAYindex}{16}
\affiliation{\ORSAY}
\newcommand*{\INFNFE}{INFN, Sezione di Ferrara, 44100 Ferrara, Italy}
\newcommand*{\INFNFEindex}{17}
\affiliation{\INFNFE}
\newcommand*{\INFNFR}{INFN, Laboratori Nazionali di Frascati, 00044 Frascati, Italy}
\newcommand*{\INFNFRindex}{18}
\affiliation{\INFNFR}
\newcommand*{\INFNGE}{INFN, Sezione di Genova, 16146 Genova, Italy}
\newcommand*{\INFNGEindex}{19}
\affiliation{\INFNGE}
\newcommand*{\INFNRO}{INFN, Sezione di Roma Tor Vergata, 00133 Rome, Italy}
\newcommand*{\INFNROindex}{20}
\affiliation{\INFNRO}
\newcommand*{\INFNTUR}{INFN, Sezione di Torino, 10125 Torino, Italy}
\newcommand*{\INFNTURindex}{21}
\affiliation{\INFNTUR}
\newcommand*{\INFNCAT}{INFN, Sezione di Catania, 95123 Catania, Italy}
\newcommand*{\INFNCATindex}{22}
\affiliation{\INFNCAT}
\newcommand*{\INFNPAV}{INFN, Sezione di Pavia, 27100 Pavia, Italy}
\newcommand*{\INFNPAVindex}{23}
\affiliation{\INFNPAV}
\newcommand*{\JMU}{James Madison University, Harrisonburg, Virginia 22807, USA}
\newcommand*{\JMUindex}{24}
\affiliation{\JMU}
\newcommand*{\KNU}{Kyungpook National University, Daegu 41566, Republic of Korea}
\newcommand*{\KNUindex}{25}
\affiliation{\KNU}
\newcommand*{\LAMAR}{Lamar University, Beaumont, Texas 77710, USA}
\newcommand*{\LAMARindex}{26}
\affiliation{\LAMAR}
\newcommand*{\MIT}{Massachusetts Institute of Technology, Cambridge, Massachusetts  02139, USA}
\newcommand*{\MITindex}{27}
\affiliation{\MIT}
\newcommand*{\MISS}{Mississippi State University, Mississippi State, Mississippi 39762, USA}
\newcommand*{\MISSindex}{28}
\affiliation{\MISS}
\newcommand*{\UNH}{University of New Hampshire, Durham, New Hampshire 03824, USA}
\newcommand*{\UNHindex}{29}
\affiliation{\UNH}
\newcommand*{\NMSU}{New Mexico State University, Las Cruces, New Mexico 88003, USA}
\newcommand*{\NMSUindex}{30}
\affiliation{\NMSU}
\newcommand*{\NSU}{Norfolk State University, Norfolk, Virginia 23504, USA}
\newcommand*{\NSUindex}{31}
\affiliation{\NSU}
\newcommand*{\OHIOU}{Ohio University, Athens, Ohio 45701, USA}
\newcommand*{\OHIOUindex}{32}
\affiliation{\OHIOU}
\newcommand*{\ODU}{Old Dominion University, Norfolk, Virginia 23529, USA}
\newcommand*{\ODUindex}{33}
\affiliation{\ODU}
\newcommand*{\JLUGiessen}{II Physikalisches Institut der Universitaet Giessen, 35392 Giessen, Germany}
\newcommand*{\JLUGiessenindex}{34}
\affiliation{\JLUGiessen}
\newcommand*{\URICH}{University of Richmond, Richmond, Virginia 23173, USA}
\newcommand*{\URICHindex}{35}
\affiliation{\URICH}
\newcommand*{\ROMAII}{Universit\`{a} di Roma Tor Vergata, 00133 Rome, Italy}
\newcommand*{\ROMAIIindex}{36}
\affiliation{\ROMAII}
\newcommand*{\SDU}{Shandong University, Qingdao, Shandong 266237, China}
\newcommand*{\SDUindex}{37}
\affiliation{\SDU}
\newcommand*{\MSU}{Skobeltsyn Institute of Nuclear Physics, Lomonosov Moscow State University, 119234 Moscow, Russia}
\newcommand*{\MSUindex}{38}
\affiliation{\MSU}
\newcommand*{\SCAROLINA}{University of South Carolina, Columbia, South Carolina 29208, USA}
\newcommand*{\SCAROLINAindex}{39}
\affiliation{\SCAROLINA}
\newcommand*{\TEMPLE}{Temple University, Philadelphia, Pennsylvania 19122, USA}
\newcommand*{\TEMPLEindex}{40}
\affiliation{\TEMPLE}
\newcommand*{\JLAB}{Thomas Jefferson National Accelerator Facility, Newport News, Virginia 23606, USA}
\newcommand*{\JLABindex}{41}
\affiliation{\JLAB}
\newcommand*{\ULS}{Universidad de La Serena, Avda Juan Cisternos 1200, La Serena, Chile}
\newcommand*{\ULSindex}{42}
\affiliation{\ULS}
\newcommand*{\UTFSM}{Universidad T\'{e}cnica Federico Santa Mar\'{i}a, Casilla 110-V, Valpara\'{i}so, Chile}
\newcommand*{\UTFSMindex}{43}
\affiliation{\UTFSM}
\newcommand*{\INSUBRIA}{Universit\`{a} degli Studi dell'Insubria, 22100 Como, Italy}
\newcommand*{\INSUBRIAindex}{44}
\affiliation{\INSUBRIA}
\newcommand*{\BRESCIA}{Universit\`{a} degli Studi di Brescia, 25123 Brescia, Italy}
\newcommand*{\BRESCIAindex}{45}
\affiliation{\BRESCIA}
\newcommand*{\GLASGOW}{University of Glasgow, Glasgow G12 8QQ, United Kingdom}
\newcommand*{\GLASGOWindex}{46}
\affiliation{\GLASGOW}
\newcommand*{\UTK}{University of Tennessee, Knoxville, Tennessee 37996, USA}
\newcommand*{\UTKindex}{47}
\affiliation{\UTK}
\newcommand*{\YORK}{University of York, York YO10 5DD, United Kingdom}
\newcommand*{\YORKindex}{48}
\affiliation{\YORK}
\newcommand*{\TELAVIV}{University of Tel Aviv, Tel Aviv 6997801, Israel}
\newcommand*{\TELAVIVindex}{49}
\affiliation{\TELAVIV}
\newcommand*{\VIRGINIA}{University of Virginia, Charlottesville, Virginia 22901, USA}
\newcommand*{\VIRGINIAindex}{50}
\affiliation{\VIRGINIA}
\newcommand*{\WM}{College of William and Mary, Williamsburg, Virginia 23187, USA}
\newcommand*{\WMindex}{51}
\affiliation{\WM}
\newcommand*{\YEREVAN}{Yerevan Physics Institute, 375036 Yerevan, Armenia}
\newcommand*{\YEREVANindex}{52}
\affiliation{\YEREVAN} 


\newcommand*{\NOWDOTA}{DOTA, ONERA, Universit\'{e} Paris-Saclay, 91120, Palaiseau, France}


\author{P.~Chatagnon}
\email{Corresponding author: pierre.chatagnon@cea.fr}
\affiliation{\SACLAY}
\author{V.~Kubarovsky}
\affiliation{\JLAB}
\author {R.~Paremuzyan} 
\affiliation{\JLAB}
\affiliation{\UNH}
\author{S.~Stepanyan}
\affiliation{\JLAB}
\author{M.~Tenorio}
\affiliation{\ODU}
\author{R.~Tyson}
\affiliation{\GLASGOW}


\author {A.G.~Acar} 
\affiliation{\YORK}
\author {P.~Achenbach} 
\affiliation{\CNU}
\author {J.S.~Alvarado} 
\affiliation{\ORSAY}
\author {M.J.~Amaryan} 
\affiliation{\ODU}
% ...

\collaboration{The CLAS Collaboration}
\noaffiliation

\date{\today}
\begin{abstract}
We present measurements of the total and differential cross sections for near-threshold J/$\psi$ photoproduction obtained with the CLAS12 detector at the Thomas Jefferson National Accelerator Facility. The results are based on data collected during the Fall 2018 and Spring 2019 running periods, using electron beams with energies of 10.6 and 10.2 GeV, respectively, scattered off a liquid-hydrogen target.
Near-threshold J$/\psi$ photoproduction offers a unique sensitivity to the strong interaction in the non-perturbative regime of Quantum Chromodynamics (QCD). The energy dependence of the cross section constrains the underlying J$/\psi$ production mechanisms, including multi-gluon exchange and potential baryonic excitations. Additionally, the $t$-dependence of the differential cross section can be related to the transverse spatial distribution of gluons in the proton, providing critical input for theoretical descriptions of the gluonic structure of the proton. An interpretation of the results in terms of the gluon content of the proton is presented, providing new experimental constraints on QCD-inspired models of the proton structure and the role of gluonic degrees of freedom in hadronic mass generation.
\end{abstract}
\maketitle

\section{Introduction}
\begin{figure}[hbtp]
\centering
\includegraphics[width=\linewidth]{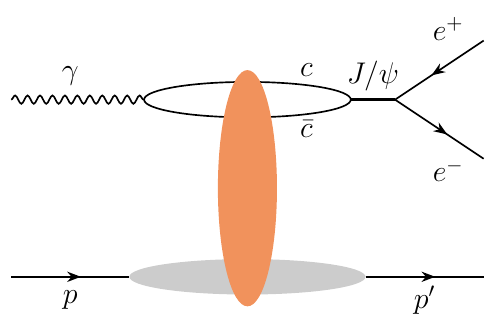}
\caption{(Color online) Diagram of exclusive J$/\psi$ photoproduction off a proton, with the J$/\psi$ decaying into an electron–positron pair. The orange ellipse represents the colorless gluons exchange in the $t$-channel, as assumed in several models describing the process near threshold~\cite{PhysRevD.100.014032,PhysRevD.103.096010,PhysRevD.104.054015,Mamo:2022eui,Guo:2023pqw}.}
\label{fig:diagram_jpsi}
\end{figure}

The photoproduction of heavy vector mesons has long been identified as an important process to probe the gluon content of nucleons. Although the exclusive photoproduction of J$/\psi$ on a proton, represented in Fig.~\ref{fig:diagram_jpsi}, has already been measured in experiments at Cornell~\cite{PhysRevLett.35.1616}, SLAC~\cite{PhysRevLett.35.483}, HERA~\cite{Zeus:2002fa,H1:2005dtp,H1:2013okq}, and at the LHC in ultra-peripheral collisions~\cite{ALICE:2014eof, ALICE:2018oyo, LHCb:2018rcm}, such a measurement near the production threshold only became possible with the 12-GeV upgrade of the CEBAF accelerator~\cite{Adderley:2024czm} at the Thomas Jefferson National Accelerator Facility (JLab). The first two of the four recent measurements at JLab were performed by the GlueX Collaboration~\cite{GlueX:2019mkq, PhysRevC.108.025201} with a tagged-photon beam incident on a liquid-hydrogen target. Both the total cross section as a function of the energy of the incoming real photon and the differential cross section as a function of the squared momentum transfer $t$, were extracted. An interpretation of these data in terms of the J$/\psi$-proton scattering length was subsequently provided in Ref.~\cite{Strakovsky:2019bev}. The third measurement was carried out by the E12-16-007 experiment in Hall C (known as the J$/\psi$-007 experiment) using an untagged photon beam scattering off a proton target. This experiment reported the differential cross section as a function of $t$ for the electron-positron final state in Ref.~\cite{Duran2023}. The J$/\psi$-007 experiment also provided an interpretation of their results in terms of gluon Gravitational Form Factors (GFFs) of the proton~\cite{Kobzarev:1962wt,Pagels:1966zza,Ji:1996ek,Burkert:2018bqq, Polyakov:2018zvc, Lorce:2018egm, Burkert:2023wzr}. Recently, the J$/\psi$-007 Collaboration reported new measurements of the total and differential cross sections for the di-muon final state in a preprint~\cite{007:2026dow}. 

Various theoretical models have been developed to relate the differential cross section for near-threshold J$/\psi$ photoproduction to the gluon content of the proton by describing the reaction in terms of gluon exchange between the proton and the J$/\psi$. Recent theoretical developments~\cite{PhysRevD.100.014032,PhysRevD.103.096010,PhysRevD.104.054015,Mamo:2022eui,Guo:2023pqw} have suggested that the gluon GFFs of the proton can be accessed via the measurement of the $t$-dependence of the cross section. 
Defined through matrix elements of the nucleon energy-momentum tensor, four gluon GFFs $A_g(t)$, $B_g(t)$, $C_g(t)$, and $\bar{C}_g(t)$ encode fundamental properties of the nucleon such as its mass and angular momentum distributions, as well as the internal distributions of pressure and shear forces. Lattice QCD calculations have also recently provided reliable estimates for the gluon GFFs~\cite{Shanahan:2018pib,Pefkou:2021fni,Hackett:2023rif}. 

Models of J$/\psi$ photoproduction based on holographic-QCD have been developed in Refs.~\cite{Mamo:2022eui,PhysRevD.101.086003,PhysRevD.103.094010,PhysRevD.104.066023}. Several approaches based on the Generalized Parton Distribution (GPD) formalism have been developed to describe the exclusive production of J$/\psi$ in the diffractive regime at large energies~\cite{Radyushkin:1996ru, Collins:1996fb, Hoodbhoy:1996zg, Frankfurt:1997ha}. GPD models at leading-order, where GFFs arise as moments of the gluon GPDs of the proton, have also been proposed to describe the process near threshold in Refs.~\cite{PhysRevD.103.096010,Guo:2023pqw}. More recently, a GPD model has been computed at next-to-leading order~\cite{Guo:2025jiz}, further strengthening the validity of such models at energies near the J$/\psi$ production threshold. Finally, a model solely based on Pomeron exchange, and thus unrelated to the gluon dynamics in the proton, has been proposed in Ref.~\cite{Tang:2025qqe} to describe the exclusive photoproduction of vector mesons, including J$/\psi$, from their production thresholds to large incoming photon energies. 

The interpretation of the differential cross section in terms of gluon GFFs is valid only if the process is indeed dominated by two-gluon exchange. A significant contribution of other reaction mechanisms could prohibit such an interpretation. In particular, the impact of open-charm intermediate states~\cite{Du:2020bqj, PhysRevD.108.054018} and possible pentaquark contributions~\cite{Kubarovsky:2015aaa,Guo:2015umn,Eides:2015dtr,Blin:2016dlf,Strakovsky:2023kqu} needs to be understood to properly describe the J$/\psi$ photoproduction process near threshold. Open-charm box diagrams can generate cusp-like structures in the cross section near the $\Lambda_c\overline{D}$ and $\Lambda_c\overline{D}^*$ thresholds. Such effects may be compatible with features observed in the GlueX total cross-section data at photon energies around 9~GeV~\cite{PhysRevC.108.025201,Du:2020bqj}.

In this context, new experimental input on near-threshold J/$\psi$ photoproduction is essential to clarify the role of the various reaction mechanisms that may contribute in this energy regime. In this work, we present measurements of the total and differential cross sections for J/$\psi$ photoproduction near threshold obtained with the CLAS12 detector at JLab. The kinematic coverage of CLAS12 enables a simultaneous investigation of the energy and $t$ dependencies of the reaction, providing complementary information to existing measurements from the GlueX and J$/\psi$-007 Collaborations.

In \Cref{sec:CLAS12}, the CLAS12 experiment is described, and the characteristics of the subsystems relevant to these measurements are detailed. \Cref{sec:Analysis} outlines the various steps of the analysis, including particle-identification algorithms, Monte Carlo simulations, and the estimation of the incoherent background using a reweighting method based on Boosted Decision Trees applied to mixed-event samples. The extraction of the cross sections is discussed in \Cref{sec:CS}, while systematic uncertainties are addressed in \Cref{sec:Syst}. The final results and their interpretation are presented in \Cref{sec:Results}. Finally, all data points obtained in this analysis are provided in \Cref{Appendix_Int_Res}, and are also available in the CLAS Physics Database~\cite{database}.

\section{The CLAS12 detector}
\label{sec:CLAS12}

The data analyzed in this work were collected with the CLAS12 detector~\cite{BURKERT2020163419}, located in Experimental Hall~B at JLab. CLAS12 is a large-acceptance spectrometer constructed as part of the 12-GeV upgrade of the CEBAF accelerator. Its design provides broad kinematic coverage, making it particularly well suited for measurements of exclusive reactions.
The acceptance of CLAS12 is divided into two main regions. The Central Detector (CD) surrounds the target within a 5-T superconducting solenoid magnet and provides polar-angle coverage from $35^\circ$ to $125^\circ$. The Forward Detector (FD) is built around a six-coil superconducting torus magnet~\cite{Fair:2020yfx} and covers polar angles from $5^\circ$ to $35^\circ$.
The continuous electron beam delivered by the CEBAF accelerator was incident on a 5-cm-long liquid-hydrogen target positioned at the center of the CD.

Only the FD was used in this analysis. In this region, charged-particle tracks were reconstructed by the Drift Chamber (DC) system~\cite{MESTAYER2020163518}, which provided a relative momentum resolution better than 1\% for the datasets considered.
Electron identification in the FD was achieved using the High Threshold Cherenkov Counter (HTCC)~\cite{Sharabian:2020whm}, which separated leptons from charged pions for momenta below 4.9~GeV, in combination with the forward Electromagnetic Calorimeter (ECAL)~\cite{Asryan:2020iqj}. The ECAL is segmented longitudinally into three layers: the Pre-shower Calorimeter (PCAL), the Inner Electromagnetic Calorimeter (ECin), and the Outer Electromagnetic Calorimeter (ECout).
The Forward Time-of-Flight (FTOF) system~\cite{Carman:2020fsv} provides charged-particle timing in the FD and was used to identify protons through measurements of their velocity.

The datasets used in this analysis were collected during the Fall 2018 and Spring 2019 running periods, also referred as Run Group A (RG-A). The experimental setup was identical for both runs, with the exception of the electron beam energy, which was 10.6~GeV for Fall 2018 and 10.2~GeV for Spring 2019. In addition, the Fall 2018 data were acquired with two distinct torus magnet configurations.
The Spring 2019 run and the first part of the Fall 2018 run were taken with the torus magnet set to the inbending configuration at its nominal maximum field strength, corresponding to an integrated magnetic field $\int\! B\,\mathrm{d}\ell$ ranging from approximately 2.8~Tm at 5$^\circ$ to 0.54~Tm at 40$^\circ$. In this configuration, negatively charged particles were bent toward the beamline. The second part of the Fall 2018 dataset was collected with the opposite torus polarity, referred to as the outbending configuration.

The electron trigger used during data collection required pre-defined DC track candidates matched to a cluster in the HTCC with a threshold exceeding two photoelectrons, a minimum deposited energy of 300~MeV in ECAL, and a minimum energy deposition in multiple ECAL layers within a given sector. This trigger configuration provided an efficiency exceeding 99\%~\cite{RAYDO2020163529}.
During these run periods, the CLAS12 detector and its data acquisition system~\cite{Boyarinov:2020yry} operated at luminosities of the order of $10^{35}$~cm$^{-2}$s$^{-1}$. The beam current delivered by the CEBAF accelerator ranged from 40 to 55~nA, resulting in typical trigger rates of approximately 15~kHz and data acquisition rates of about 500~MB/s of raw data, with an overall livetime exceeding 90\%.


\section{Analysis procedure}
\label{sec:Analysis}
This analysis measured the cross section of J$/\psi$ photoproduction and, since CLAS12 does not operate with a dedicated photon beam, the reaction was identified in the photoproduction regime where the beam electron emits a nearly real photon that interacts with the target proton producing a J$/\psi$ as
\begin{equation}
    ep \rightarrow (e')\gamma p \rightarrow (e') {\rm J} /\psi~p' \rightarrow (X) e^+ e^- p'.
\end{equation}
Most of the time, the scattered electron was only slightly deflected at very small angles and therefore fell outside the CLAS12 acceptance. Nevertheless, its four-vector, $p_X$, could be fully reconstructed from the four-vectors of all detected final-state particles ($p_{e^+}$, $p_{e^-}$, and $p_{p'}$), and of the incoming beam electron and target proton ($p_{b}$ and $p_{p}$), as
\begin{equation}
\label{eqn:reaction}
    p_X = p_{b}+p_{p}-p_{e^+}-p_{e^-}-p_{p'}.
\end{equation}

Hence, the measurement of exclusive J$/\psi$ photoproduction with CLAS12 required the identification of the recoil proton and the two leptons originating from the J$/\psi$ decay. The following sections describe in detail the procedures used to identify and reconstruct this final state.

\subsection{Particle Identification}
An initial particle identification was provided by the CLAS12 event builder~\cite{ZIEGLER2020163472}, which combines information from various subsystems. Electrons and positrons were identified by matching a reconstructed track in the DC with a cluster in the ECAL and a signal in the HTCC corresponding to at least two photoelectrons. Candidate lepton clusters were required to have an energy deposition in the PCAL that exceeded 60~MeV and a sampling fraction, defined as the ratio of the total energy deposited in the calorimeter to the reconstructed particle momentum, within five standard deviations of the expected value for electrons.
To ensure reliable energy reconstruction, clusters located near the edges of the ECAL were excluded, as were regions of the calorimeter known to have reduced efficiency during data collection. These fiducial cuts were applied consistently to both the experimental data and the Monte Carlo (MC) simulations throughout the analysis to ensure consistent lepton reconstruction efficiencies. 
The two decay leptons were also required to originate from the same beam bunch. This was enforced by requiring their reconstructed times to differ by less than 4~ns, corresponding to the time interval between consecutive CEBAF beam bunches.

Protons were identified by reconstructing the speed of positively charged tracks using time-of-flight and path-length measurements and comparing it with the expected speed calculated from the track momentum under the proton mass hypothesis. Finally, the momenta of the protons were corrected for energy loss in the detector materials using MC simulations. Lepton momentum corrections were obtained using a data-driven approach based on radiative elastic events and were applied independently for each polarity configuration of the torus magnet.

\subsection{BDT-Based Lepton Identification}

A large contamination of positively and negatively charged pions was observed in the positron and electron samples at momenta above approximately $4.9$ GeV. Above this momentum threshold, the HTCC lacked reliable discrimination power between leptons and charged pions, and only the ECAL could be used to distinguish them~\cite{Sharabian:2020whm}. To complement the sampling fraction-based method used in the CLAS12 event builder, a Boosted Decision Tree (BDT) algorithm was developed. This algorithm, largely based on the one used in the published CLAS12 Timelike Compton Scattering analysis~\cite{PhysRevLett.127.262501}, used the reconstructed electromagnetic shower parameters measured by the ECAL to distinguish between leptons and charged pions.

The algorithm, implemented with the TMVA package~\cite{TMVA:2007ngy, Voss:2007jxm}, was trained on single-particle MC events. Signal events were defined as correctly identified positrons and electrons, while background events were defined as charged pions that were misidentified as leptons. Several sets of input variables were tested. For the present analysis, nine input variables were provided to the BDT: the polar and azimuthal angles, the momentum in the lab frame, the sampling fraction, and the logarithmic energy-weighted radius of the reconstructed energy deposition in each of the three calorimeter layers. Six different BDTs were trained: one for electrons and one for positrons, for each run period, accounting for changes in the torus magnet configuration. Cuts on BDT classifiers 
were optimized to suppress 
pion contamination in the electron and positron samples by approximately an order of magnitude, while maintaining a lepton detection efficiency above 95\%.

Since the BDT classifiers were trained on MC samples, additional validation was performed on real data. Specific reactions were used to provide pure samples of signal and background events.
For the signal, we used radiative events where electrons or positrons emitted a bremsstrahlung photon while traversing the target and detector material, and that was detected in the ECAL,
\begin{equation}
    ep\rightarrow e^{-(+)}\gamma X.
\end{equation}
These events were identified in the analysis of the polar-angle difference between the lepton and the photon, since they should be equal, as photons are emitted in the direction of the lepton. For the background, we used exclusive $\pi^+$ electroproduction events in which the pion was identified as a positron,
\begin{equation}
    ep\rightarrow e'e^+_\pi (n), 
\end{equation}
where the subscript indicates that the positron candidate was assumed to be a $\pi^+$. These events were selected by examining the missing mass of the $e' e^+$ system, calculated by assigning the pion mass hypothesis to the positron candidate. 

\begin{figure}[hbtp]
    \centering
    \includegraphics[width=\linewidth]{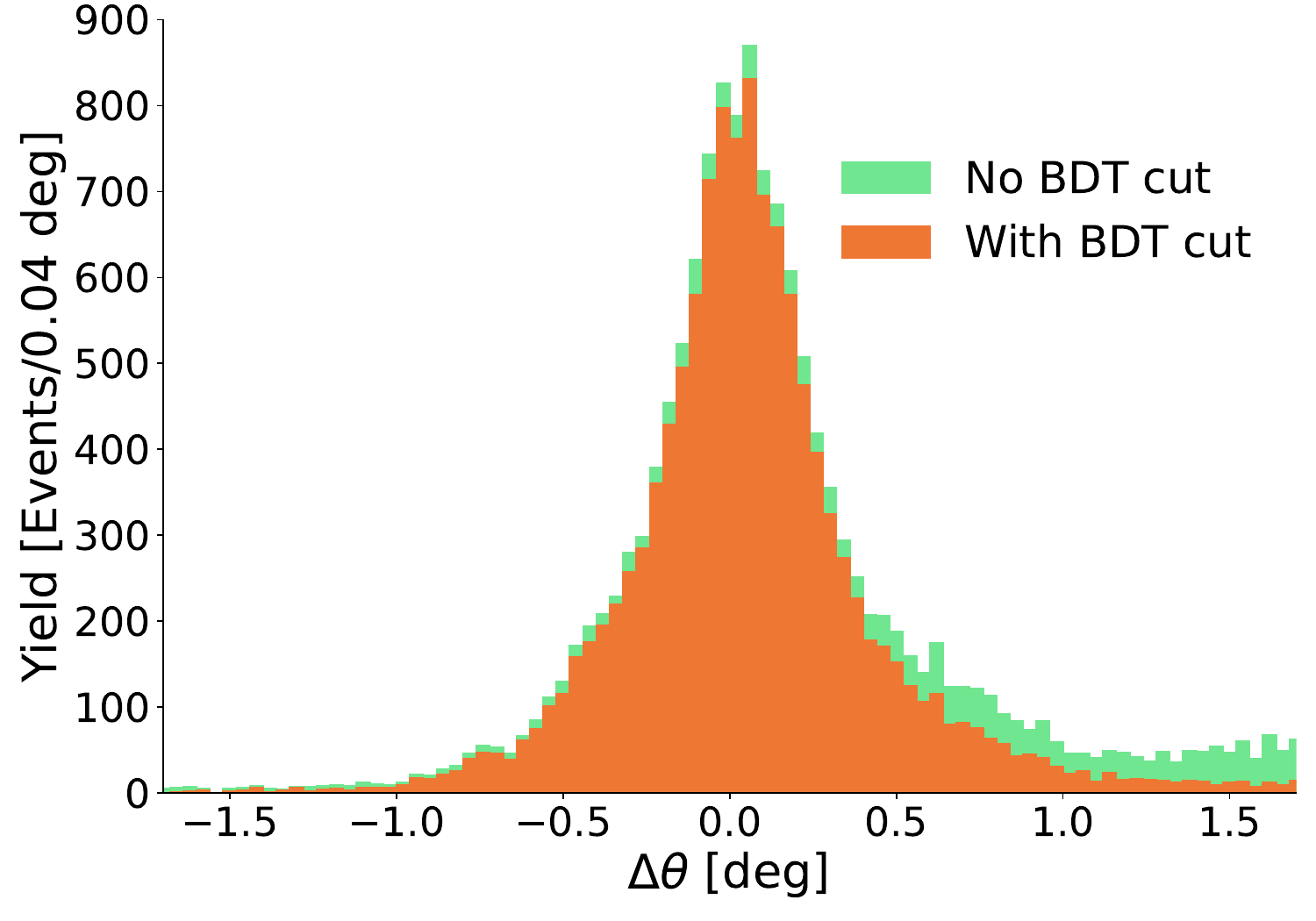}\\
    \includegraphics[width=\linewidth]{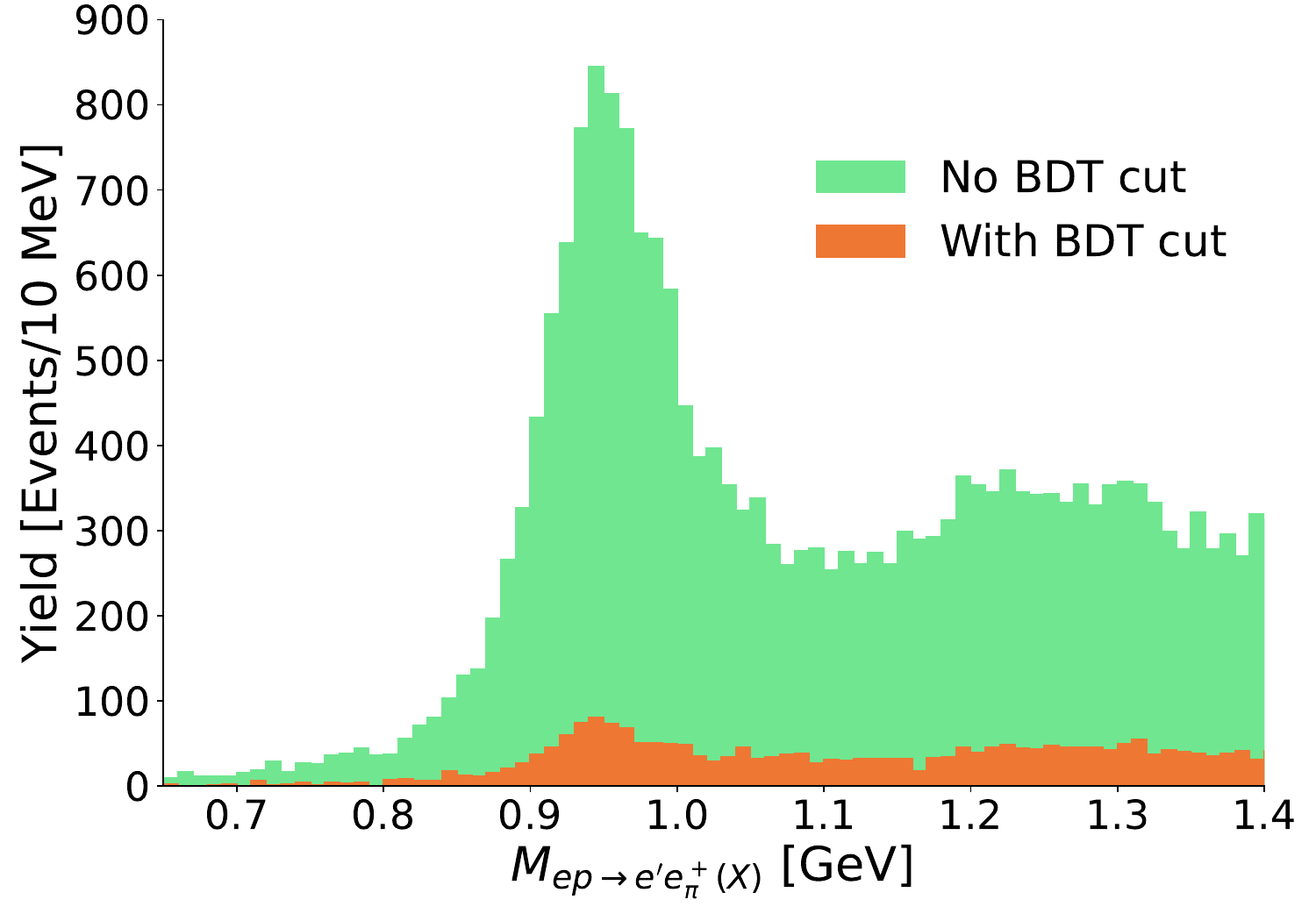}
    \caption{(Color online) Electron-photon angular correlation for radiative events (top) and missing mass spectrum of misidentified positron events (bottom). The green spectra were obtained without any additional BDT-based lepton identification. The orange spectra were obtained after the selection cut was applied on the output of the BDT-based lepton identification algorithm.}
        \label{fig:BDT_signal_background}
\end{figure}

The top panel of \Cref{fig:BDT_signal_background} shows the difference in polar angle ($\Delta\theta$) between the electron and the photon for radiative events, exhibiting a clear peak at zero. The bottom panel shows the missing mass distribution for the $e' e^+$ final state, with the $\pi^+$ mass assigned to the positron. A distinct peak is visible at the neutron mass, indicating exclusive $\pi^+$ production where the pion was identified as a positron. Both distributions are shown with standard particle identification cuts (green) and with BDT-based selections applied to the electron and positron candidates (orange). The lepton identification efficiency was extracted from the number of events in the Gaussian peak at $\Delta\theta=0^\circ$, before and after the BDT cuts. The background rejection was obtained by fitting the neutron peak in the missing mass distribution, before and after the BDT cuts.

The signal efficiency and background rejection of this lepton identification algorithm differed a priori between MC and real data. To account quantitatively for this discrepancy, especially in the calculation of the efficiency-correction factor in \Cref{sec:eff_corr_factor}, the signal efficiency ratio was calculated as a function of the cut applied to the BDT outputs. At the standard output threshold value, the efficiency ratio between MC and data deviates by 5\% for both electrons and positrons.

\subsection{Event Selection and Kinematic Reconstruction}
An initial event sample was defined by selecting events with exactly one proton, exactly one positron, and exactly one electron, all reconstructed in the FD, with both leptons required to have momenta greater than 1.7~GeV. Photoproduction events, in which the incoming beam electron emits a photon that subsequently interacts with a target proton, were then isolated using exclusivity cuts. Two exclusivity variables were considered: the squared missing mass of the undetected system $\MM2$ and the estimated virtuality of the incoming photon $\Qt2$, calculated under the assumption of a single undetected massless scattered electron. These quantities were reconstructed as
\begin{equation}
    \MM2 = p^2_X,
\end{equation}
and 
\begin{equation}
    \Qt2 = 2 E_{b} E_X (1-\cos \theta_X),
\end{equation}
where $p_X$ is defined in \Cref{eqn:reaction}, $E_X$ is the energy, and $\theta_X$ is the polar angle calculated for the undetected scattered electron in the laboratory frame.

\begin{figure}[hbtp]
    \centering
    \includegraphics[width=\linewidth]{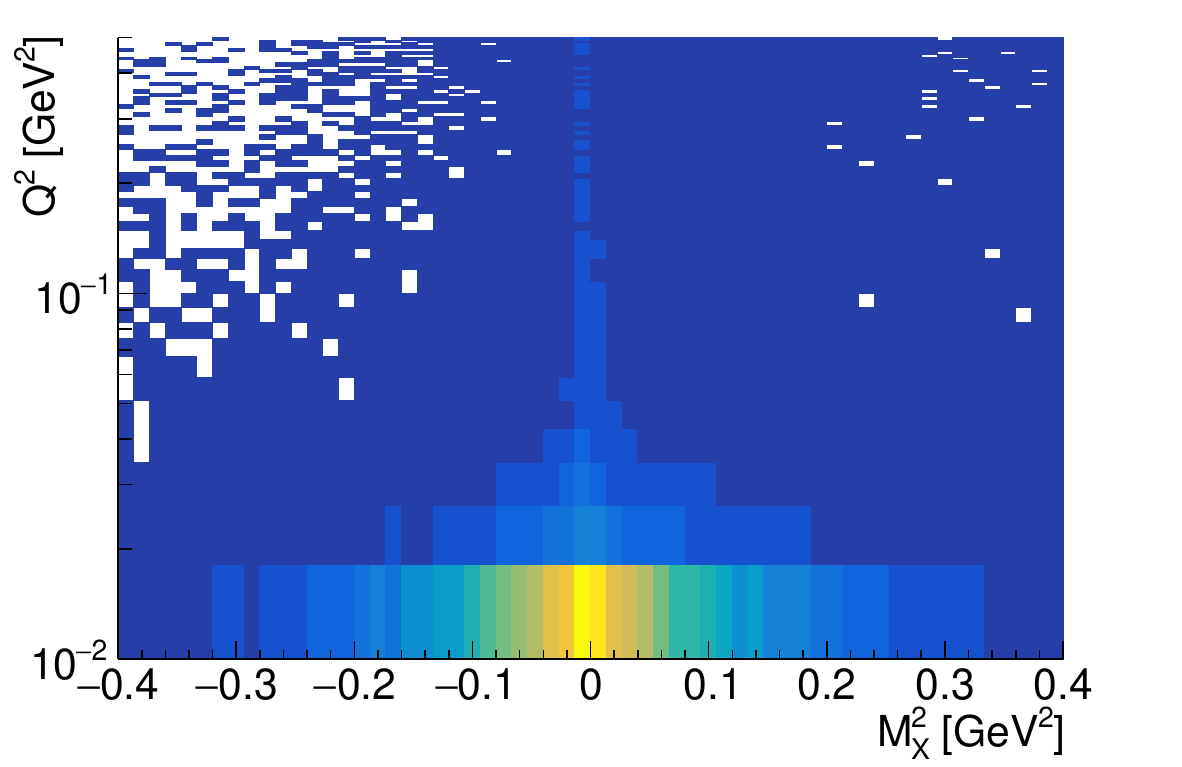}
    \caption{(Color online) Reconstructed incoming photon virtuality $\Qt2$ as a function of the squared missing mass $\MM2$ for selected photoproduction events in the Fall 2018 inbending dataset.}
    \label{fig:Exclusivity_Data}
\end{figure}

For this analysis, the exclusivity variables were restricted to $\Qt2 < 0.5~\mathrm{GeV}^2$ and $|\MM2| < 0.4~\mathrm{GeV}^2$. Figure~\ref{fig:Exclusivity_Data} shows $\Qt2$ as a function of $\MM2$ for the Fall~2018 inbending dataset after all particle identification and exclusivity cuts were applied. A clear enhancement is observed in the region where both $\Qt2$ and $\MM2$ are small, corresponding to photoproduction events.

The initial photon energy $E_\gamma$ and the squared momentum transfer $t$ were then computed as 
\begin{equation}
    E_\gamma = E_{b} - E_X,
\end{equation}
and
\begin{equation}
    t = (p_p-p_{p'})^2.
\end{equation}

\subsection{Monte Carlo Simulations}
\label{sec:MonteCarlo}

The cross-section measurements presented in this article rely extensively on Monte Carlo simulations to understand the reaction phase space, and to correct for the geometrical acceptance of CLAS12 and the intrinsic efficiencies of its subsystems. The CLAS12 experiment is fully implemented within the dedicated GEMC software~\cite{UNGARO2020163422}, a simulation framework based on Geant4~\cite{AGOSTINELLI2003250}. After simulating the passage of particles through the detector, GEMC produces digitized hit information in the active volumes using the same data format as that recorded during data taking. To account for detection inefficiencies arising from beam-related background in the various CLAS12 detector subsystems~\cite{cn-2020-005}, a background-merging procedure was applied to the GEMC output. Real events collected with a random trigger during data taking were merged event-by-event with the simulated data at the digitization level. Further corrections to ensure optimal agreement between data and simulation are described in \Cref{sec:eff_corr_factor}. Finally, simulated events with merged background were reconstructed with the same reconstruction software chain as for real data, and analyzed with the same procedure. 

To generate the J$/\psi$ signal and the underlying non-resonant lepton-pair production continuum, known as the Bethe-Heitler (BH) process, three generators were employed. The GRAPE generator~\cite{ABE2001126} was used to generate the virtual photon contribution of the BH continuum. The TCSGen generator~\cite{JeffersonLab_TCSGen}, developed at JLab and extensively used in Ref.~\cite{PhysRevLett.127.262501}, provided the real-photon contribution of the BH continuum. The J$/\psi$ signal was produced with a dedicated generator, JPsiGen~\cite{JeffersonLab_JPsiGen}. Radiative effects were included using the formalism of Ref.~\cite{Heller:2018ypa} for the BH continuum and Ref.~\cite{Ehlotzky1968} for the J$/\psi$ production.

Twenty-four different MC samples were produced to account for each experimental configuration and beam current. Additional samples without radiative effects were produced to evaluate the radiative corrections detailed in \Cref{sec:RadiativeCorrection}.

\subsection{Mixed-Event Background}
\label{sec:MixedEvent}

In addition to the BH continuum and J$/\psi$ peak, the final state of interest can be reconstructed when a misidentified positron, which was not removed by the BDT-based lepton identification or a real positron from a photon conversion, was detected in coincidence with an electron and a proton. The reconstructed photon virtuality $\Qt2$ for these events does not peak at zero, and the MC samples described previously do not contain these events. To model this incoherent background, an event-mixing approach was used. An initial mixed-event sample was built by combining an electron, a positron, and a proton from three different events, and then reweighted following Ref.~\cite{Rogozhnikov:2016bdp} to better reproduce the observed data spectra.

Two samples were used for the reweighting procedure: the source sample, consisting of the mixed events, and the target sample, composed of real-data photoproduction events within the training region, defined as $|M_X^2|<0.4~{\rm GeV}^2$ and $0.5~{\rm GeV}^2<\Qt2<1.5~{\rm GeV}^2$ to prevent any possible bias in the phase space of interest. The reweighting procedure consisted of training a BDT to discriminate between these two samples, and using the output to assign weights to the events in the source sample. After the procedure, the probability distribution functions of the source and target samples were equal. Because this method matched only the shape of these distributions, the reweighted mixed‑event sample must subsequently be scaled by an overall normalization factor to reproduce the event yield in the training region.

Nine variables were used as inputs to the BDT: the invariant mass of the lepton pair, the squared missing mass, the photon energy, and the momenta and polar angles of the three final-state particles. Because $\Qt2$ was used to define the signal and training region, it was not used as a training variable. Nevertheless, it was verified that the $\Qt2$ spectra in the training region were exactly the same after the reweighting procedure was applied. Finally, the validation region was defined as $1.5~{\rm GeV}^2<\Qt2<2~{\rm GeV}^2$, where the incoherent background was expected to largely dominate. In this region, the reweighted background spectra agree with the real data within 10\%. 

\begin{figure}[h!]
  \centering
   \includegraphics[width=\linewidth]{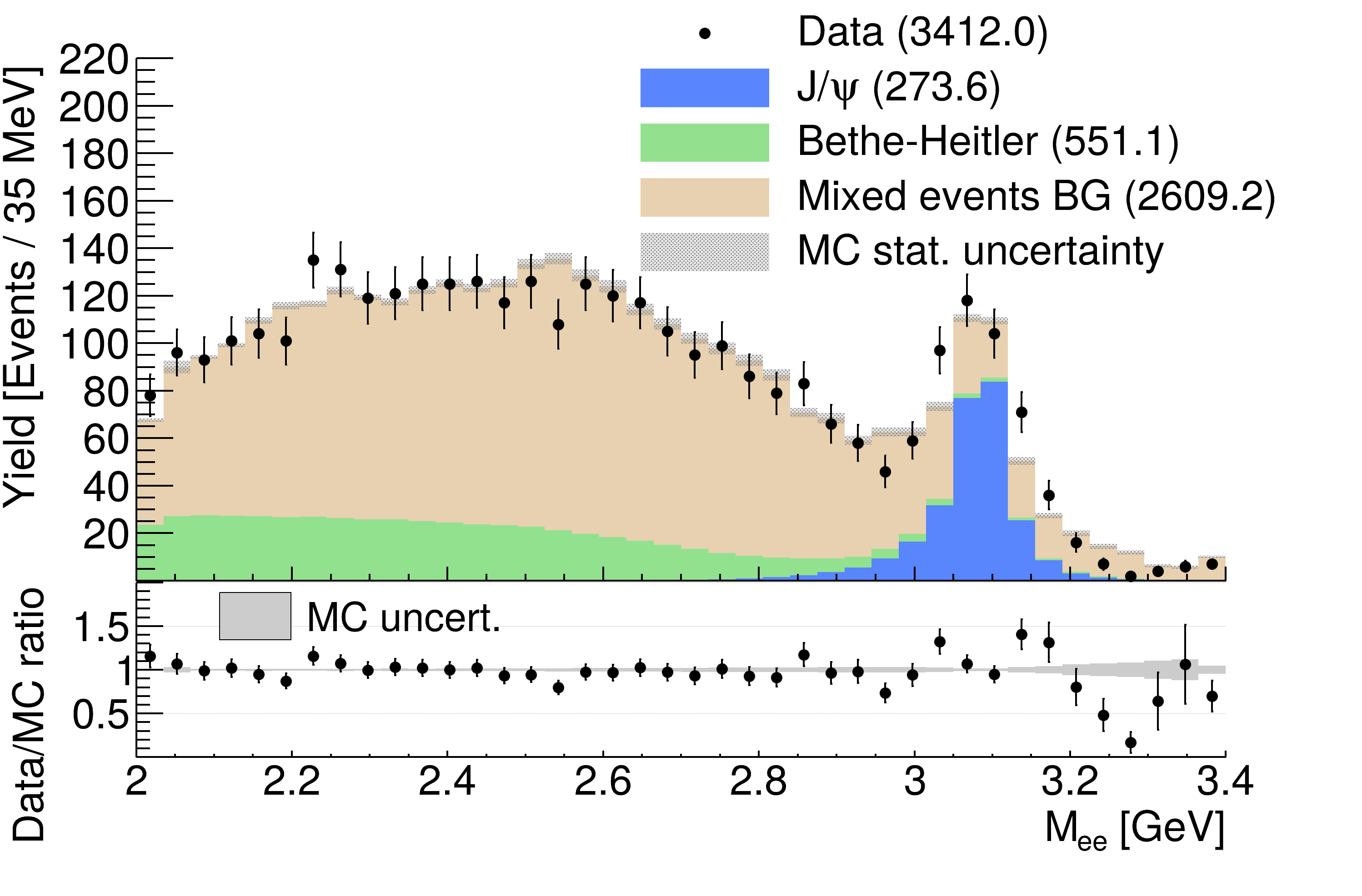}\\
   \includegraphics[width=\linewidth]{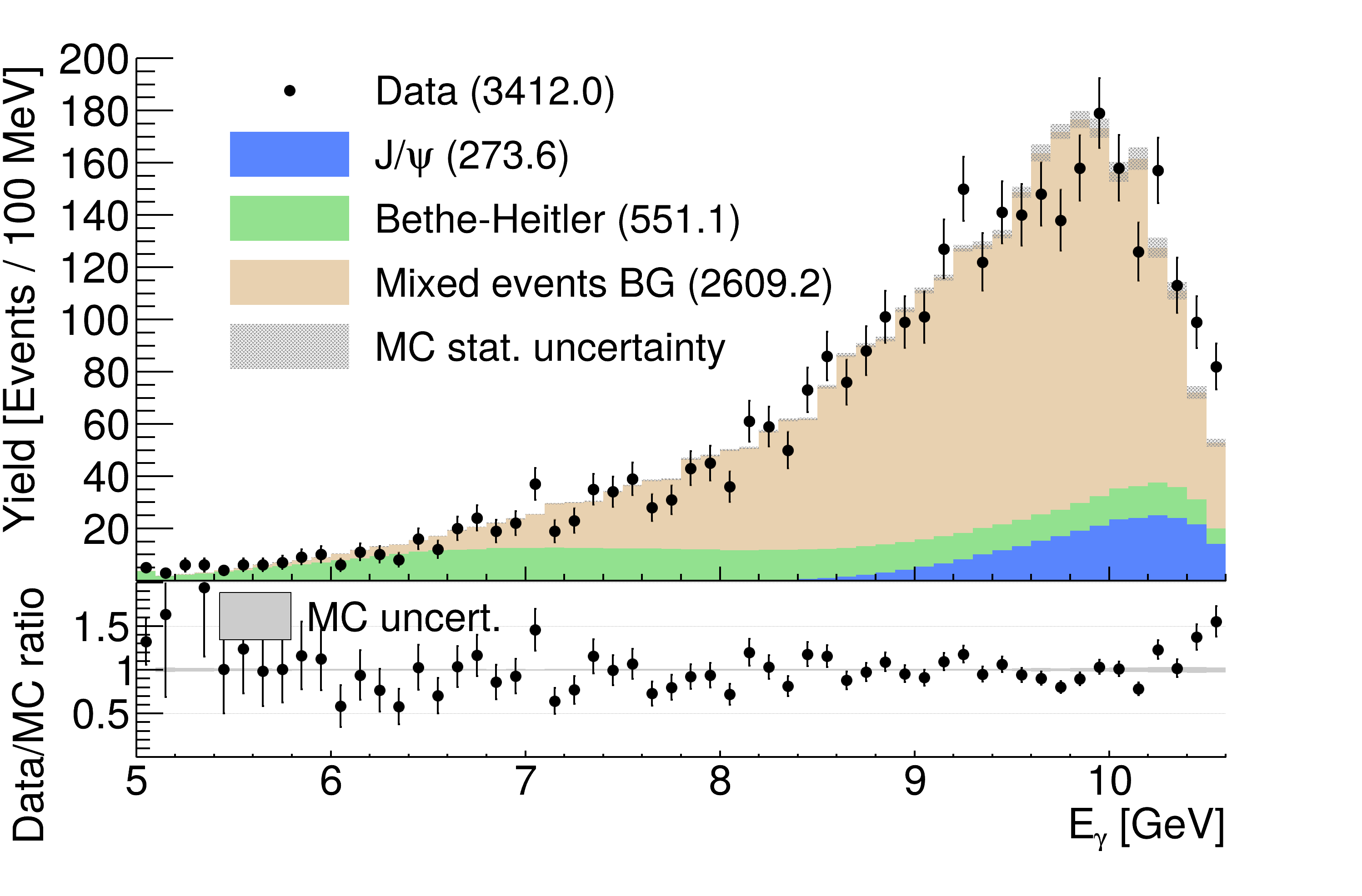}
    \caption{(Color online) Invariant mass of the electron-positron pair (top) and photon energy (bottom) for the selected photoproduction events in the Fall 2018 inbending dataset. Data points are the black circles. The MC contributions to the J$/\psi$ signal, the BH continuum, and the mixed-event background are represented using stacked histograms. The total number of events in data and for each MC contribution is indicated in parentheses in the caption of each figure. The bottom panel of each plot shows the ratio of the data to the summed MC contributions.}
    \label{fig:MCandData}
\end{figure}

For each dataset, a mixed-event sample was produced and reweighted following this procedure. Figure~\ref{fig:MCandData} shows the invariant-mass and photon-energy spectra of the Fall 2018 inbending dataset, and the associated MC samples for the BH continuum and J$/\psi$ peak. The mixed-event background is also shown. All data spectra, for each dataset, are well reproduced by the various contributions considered in the analysis. This is especially important for the computation of the efficiency-correction factor described in \Cref{sec:eff_corr_factor}.

\section{Cross-section measurement}
\label{sec:CS}

This section summarizes the steps involved in both the measurements of the total and differential cross sections of the near-threshold photoproduction of J$/\psi$. For each photon-energy bin $i$, the total cross section is expressed as
\begin{equation}
\label{eq:cs_tot}
    \sigma_i = \frac{N^{{\rm J}/\psi}_i}{ \epsilon^{\rm Det}_i    \epsilon^{\rm Rad}_i\omega_{c}} \,\frac{1}{ \mathcal{F}^\gamma_i  \mathcal{T}   B_r  },
\end{equation}
where $N^{{\rm J}/\psi}_i$ is the number of J$/\psi$ measured from data, $\epsilon^{\rm Det}_i$~is the acceptance-correction factor estimated from MC, $\epsilon^{\rm Rad}_i$~is the radiative-correction factor extracted from MC, $\omega_{c}$~is an efficiency-correction factor, $\mathcal{F}^\gamma_i$~is the accumulated-charge weighted photon flux, $\mathcal{T}$~is the target-thickness factor, and $B_r=0.0597$ is the branching ratio of the ${\rm J}/\psi\to e^+e^-$ decay~\cite{ParticleDataGroup:2024cfk}.
     
The formula for the differential cross section in a given $E_\gamma$-$t$ bin $j$ is
\begin{equation}
\label{eq:cs_diff}
    \left.\frac{d\sigma}{dt}\right\vert_j = \frac{N^{{\rm J}/\psi}_j}{   \epsilon^{\rm Det}_j   \epsilon^{\rm Rad}_j \omega_{c} } \, \frac{1}{ \mathcal{F}^\gamma_j  \mathcal{T}  B_r} \, \frac{1}{\mathcal{V}_j  \Delta t_j},
\end{equation}
where $\mathcal{V}_j$ is the bin-volume correction and $\Delta t_j = t_j^{\rm max}-t_j^{\rm min}$ is the width of the $-t$ bin. The determination of each quantity entering \Cref{eq:cs_tot,eq:cs_diff} is described below.

\subsection{J$/\psi$ Yield}

\begin{figure}[h!]
  \centering
   \includegraphics[width=\linewidth, page = 1]{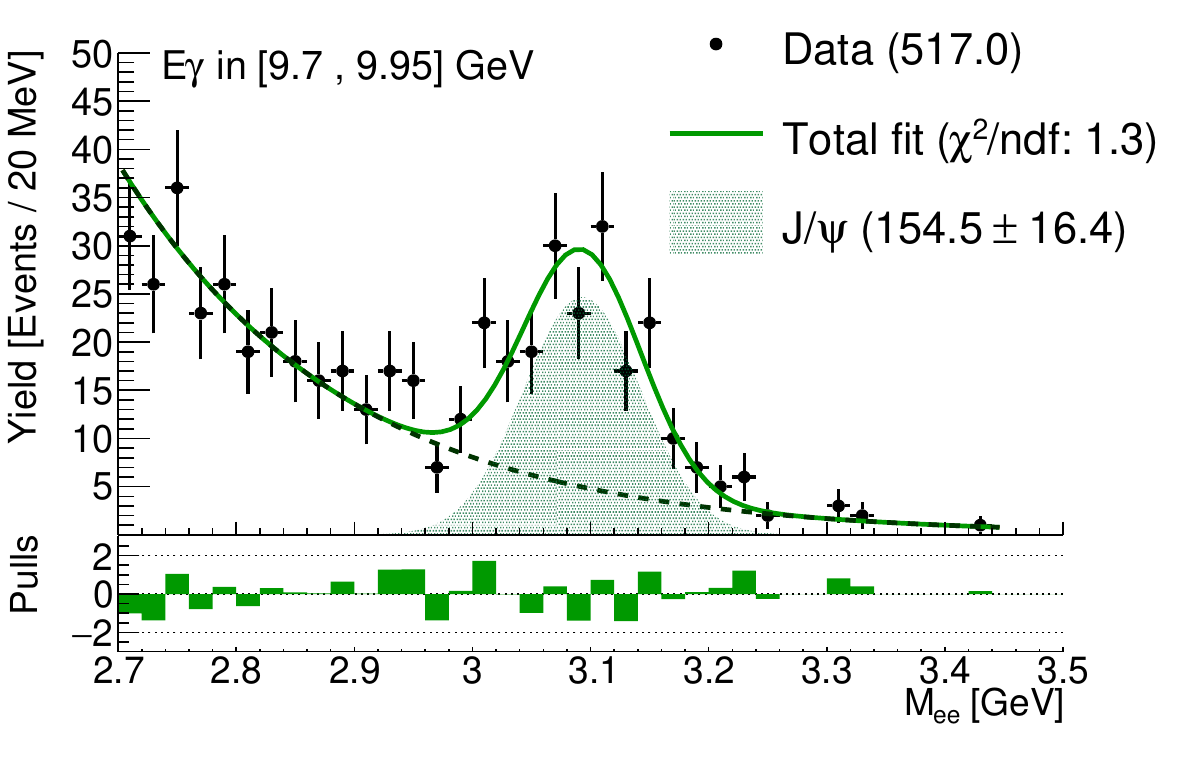}\\
   \includegraphics[width=\linewidth, page = 2]{Fit_Signal_Data_Publi.pdf}\\
    \includegraphics[width=\linewidth, page = 3]{Fit_Signal_Data_Publi.pdf}
\caption{(Color online) Invariant mass spectra of the electron-positron pair for selected photoproduction events in the J$/\psi$ mass region, combining all three datasets, and for three photon-energy bins. The J$/\psi$ yields were extracted by fitting the J$/\psi$ peak with a Gaussian signal and an exponentially decaying background, providing reasonable $\chi^2/\rm{ndf}$ (where $\rm ndf$ denotes the number of degrees of freedom of the fit). The total number of events and the J$/\psi$ yield is given in parentheses in the caption of each figure. The bottom panel of each plot shows the pulls, defined as the difference between the measured and fitted yield, normalized by the statistical uncertainty of each data point.}
\label{fig:Fit_Examples}
\end{figure}

To extract the raw J$/\psi$ yield $N^{{\rm J}/\psi}_{i/j}$, the electron-positron invariant mass spectrum, obtained from the combination of the three datasets used in this work, was fitted in the region of the J$/\psi$ peak with a Gaussian signal combined with an exponentially decaying background. Examples of such fits are shown in \Cref{fig:Fit_Examples} for three representative photon-energy bins used in this analysis. In total, 779.9$\pm$40.4 J$/\psi$ were used for these cross section measurements.

\subsection{Photon Flux}
For the selected photoproduction events, the incoming beam electron emits a high-energy photon that subsequently interacts with the target proton. These initial photons originate from two distinct mechanisms: real photons emitted by the bremsstrahlung interaction of the beam with the material of the target, and low-virtuality photons. The real-photon flux is fully described by the properties of the target material~\cite{ParticleDataGroup:2024cfk} and can be written as
\begin{equation}
     {f}(E_\gamma)_{\rm real} = \frac{l}{2X_0E_\gamma} \left(   \frac{4}{3} - \frac{4}{3} x   + x^2\right),
\end{equation} 
where $l=5~{\rm cm}$ is the length of the CLAS12 liquid-hydrogen target cell, $X_0$ is the radiation length of liquid hydrogen, and $x = E_\gamma / E_{b}$, where $E_{b}$ is the energy of the incoming electron beam. The factor of 1/2 accounts for the fact that real photons are emitted over the entire length of the target cell.

The virtual-photon flux can be expressed in the Equivalent Photon Approximation~\cite{EPA_flux_ref} as
\begin{equation}
 {f}(E_\gamma)_{\rm virtual} = \frac{\alpha}{\pi xE_{b}} \, 
\Bigl[ \bigl(1 - x + \frac{x^2}{2}\bigr) \ln \frac{Q^2_{\rm max}}{Q^2_{\rm min}} - (1 - x) \Bigr],
\end{equation}
where $\alpha$ is the fine structure constant and $Q^2_{\rm min} = m_e^2  x^2 / (1 - x)$. The $Q^2_{\rm max}$ parameter denotes the maximum possible photon virtuality and is fixed by the experimental kinematics. For this analysis, a sample of BH events simulated with the GRAPE generator was used to determine its value. Applying all analysis cuts, except the $\Qt2$ upper bound, $Q^2_{\rm max}$ was determined as the value of the generated photon virtuality below which there are the same number of events as when reconstructed using all analysis cuts, including that on $\Qt2$. Following this procedure, the limit was set at $Q^2_{\rm max}=0.1$~GeV$^2$.

Initial state radiation (ISR), where the incoming beam electron radiates energy before emitting a high-energy photon effectively reducing the incoming photon flux, was also taken into account through a correction factor. In the case of the virtual-photon flux, the correction was determined using the GRAPE generator by comparing the photon-energy spectra with and without the effect of ISR. It was found to be constant throughout the range of photon energy and equal to 0.83. For the real-photon case, the correction factor was estimated to be 0.94 following the prescription in Refs.~\cite{PhysRev.167.1280} and~\cite{PhysRev.140.B1661}, and was therefore ignored as the real-photon flux only accounts for less than a third of the total photon flux.

Finally, the current-weighted integrated flux was calculated for each photon-energy bin as
\begin{equation}
    \mathcal{F}^\gamma_{i/j} = \sum_{c} C_c   \int_{\left.{E_\gamma^{\rm min}}\right\vert_{{i/j}}}^{\left.{E_\gamma^{\rm max}}\right\vert_{{i/j}}} dE_\gamma \left( \alpha_{\rm ISR} {f}_{\rm virtual}^{c} +  {f}_{\rm real}^{c} \right), 
\end{equation}
where the sum runs over all run configurations $c$, $C_c$ is the accumulated electron-beam charge for the dataset $c$, the integral covers the photon-energy range of the bin, and $\alpha_{\rm ISR}$ is the ISR-correction factor for the virtual photon flux.

\subsection{Target-Thickness Factor}
For fixed-target experiments such as CLAS12, the integrated luminosity depends on the total beam charge incident on the target, included in the photon flux in this analysis, and on the characteristics of the target. The target-thickness factor $\mathcal{T}$ is defined as
\begin{equation}
    \mathcal{T} = \frac{ 2l \rho N_A}{e M_t},
\end{equation}
where $l$ is the length of the CLAS12 target cell, $\rho$ is the density of liquid di-hydrogen, $N_A$ is the Avogadro constant, $e$ is the elementary charge, and $M_t$ is the molar mass of di-hydrogen. The factor of 2 accounts for the two protons in each $\rm H_2$ molecule.

\subsection{Acceptance-Correction Factor}
Despite the angular coverage of CLAS12, exclusive reactions such as J$/\psi$ photoproduction have a limited geometrical acceptance, as all final-state particles must be detected to select the event. Furthermore, the intrinsic efficiencies of the individual subsystems and of the reconstruction process must be taken into account when computing the cross section. To correct for all of these effects, an acceptance-correction factor was extracted from simulation data. For each bin, all J$/\psi$ MC samples were used to randomly draw a number of events equal to $\delta$ times the number of J$/\psi$ reconstructed in data in that bin. To model the background under the J$/\psi$ peak, background events were generated according to the background fit function obtained during the J$/\psi$ yield extraction on real data, with the number of random background events set to match the integral of the fitted real-data background in that bin. These background events were then combined with the randomly selected MC J$/\psi$ events, and the resulting dataset was fitted with the same functional form applied to the real data. 
For each new random draw and fit, the acceptance-correction factor was then calculated in each bin $i$ or $j$ using the yield obtained by the MC fit $ N_{\rm MC}^{\rm Fit}$, the number of generated events after radiative effects were applied $N^{\rm GEN+RAD}_{\rm MC}$, the scaling factor $\delta$, the number of fitted events in data $N^{\rm Fit}_{\rm Data}$, and the total number of reconstructed J$/\psi$ MC events $ N^{\rm REC}_{\rm MC}$ as
\begin{equation}
    \epsilon_{i/j} = \frac{\left. N_{\rm MC}^{\rm Fit} \right\vert_{i/j}}{\left. N^{\rm GEN+RAD}_{\rm MC} \right\vert_{i/j}} \, \frac{\left. N^{\rm REC}_{\rm MC} \right\vert_{i/j} }{\delta  \left. N^{\rm Fit}_{\rm Data} \right\vert_{i/j}}.
\end{equation}
An acceptance-correction factor averaged over a thousand pseudo-experiments was then used in the cross section extraction. Various values of $\delta$ were tested, with no significant impact on the determination of this correction factor, and the value $\delta=4$ was adopted for this analysis. 

In addition, the impact of the induced polarization of the J$/\psi$ on this correction was studied. The cross section model implemented in the J$/\psi$ MC samples was reweighted according to
\begin{equation}
    {W}(\cos\theta_{\rm GJ})=1+\lambda_\theta\cos^2(\theta_{\rm GJ}),
\end{equation}
where $\theta_{\rm GJ}$ is the Gottfried-Jackson polar angle~\cite{Gottfried:1964nx} of the decay electron. The parameter $\lambda_\theta$ quantifies the angular anisotropy of the J$/\psi$ decay and is directly related to the spin-density matrix element $\rho^0_{00}$ (see Refs.~\cite{Schilling:1969um,Titov:1999eu, Faccioli:2010kd}), which represents the fraction of longitudinally polarized J$/\psi$ mesons. The impact of varying $\lambda_\theta$ is detailed in \Cref{sec:Syst}.

Typical acceptance values for J$/\psi$ photoproduction events in CLAS12 range from 2\% near threshold, up to 10\% at photon energies close to 10 GeV.

\subsection{Efficiency-Correction Factor}
\label{sec:eff_corr_factor} 

An efficiency-correction factor, $\omega_{c}$, was calculated to account for residual discrepancies between the real data and simulated efficiencies after applying the background-merging procedure described in \Cref{sec:MonteCarlo}. This factor was determined using two independent methods. The first method exploited the BH continuum in the invariant-mass range $2.4~\mathrm{GeV} < M_{ee} < 2.9~\mathrm{GeV}$, relying on the description of the real-data spectrum provided by the MC and mixed-event samples, as detailed in \Cref{sec:MonteCarlo,sec:MixedEvent}. The real-data BH yield was obtained by subtracting the incoherent background from the total number of events in this invariant-mass range. The efficiency-correction factor $\omega_{c}$ was then calculated as the ratio of the BH yield in data to the corresponding yield predicted by MC simulations.

A second determination of $\omega_{c}$ was based on the intrinsic detection efficiencies of each final-state particle. BH MC samples were used to compute the single-particle efficiency for electrons, positrons, and protons, for each value of the beam current used during data collection. In real data, the tracking efficiencies for electrons and positrons in the FD were assumed to be equal to those for charged hadrons, whereas the efficiencies of the ECAL and the HTCC were evaluated using radiative events. The difference in efficiencies of the BDT-based lepton-identification algorithm between simulation and real data was also taken into account. Finally, the ratio of real-data to simulation combined efficiencies was computed for each beam current, and a weighted average, accounting for the accumulated charge in each run period, was taken as the second estimate of the efficiency-correction factor.

Averaging the factors obtained from both methods yielded a final correction of $\omega_{c}= 0.77\pm0.08$.

\subsection{Radiative Corrections}
\label{sec:RadiativeCorrection}
A radiative-correction factor was applied to account for the shift in reconstructed kinematics that occurs when one of the leptons loses energy through radiative processes. This correction was evaluated as the ratio of events generated in a given bin $i$ or $j$ with radiative effects $\left. N^{\rm GEN+RAD}_{\rm MC} \right\vert_{i/j}$ and without radiative effects $\left. N^{\rm GEN}_{\rm MC} \right\vert_{i/j}$ as
\begin{equation}
    \epsilon^{\rm Rad}_{i/j} = \frac{\left. N^{\rm GEN+RAD}_{\rm MC} \right\vert_{i/j}}{\left. N^{\rm GEN}_{\rm MC} \right\vert_{i/j}}.
\end{equation}

The correction is larger than one at small photon energies and gradually falls below unity as the energy of the incoming photon increases. This reflects the fact that the reconstructed photon energy, estimated as the sum of energies of final state particles minus the mass of the target proton, is smaller than the actual energy as a consequence of radiative energy losses.

\subsection{Bin-Volume Correction}
The differential cross section was extracted in three bins of incident photon energy. To account for the finite bin size and the fact that certain values of $t$ are kinematically forbidden at a given photon energy, a bin‑volume correction factor, $\mathcal{V}_j$, was applied. It is defined as the fraction of the bin that lies within the kinematically allowed phase space and ensures that the extracted cross section is properly normalized to the portion of the bin that is physically accessible.

\section{Systematic uncertainties}
\label{sec:Syst}

A complete study of the systematic uncertainties on the cross-section computations was performed to assess the impact of the various selection criteria and corrections applied in this analysis. For each identified source of systematics, several variations of the complete analysis were performed. For each bin, the standard deviation of the variation in the extracted cross section was assigned as the systematic uncertainty for a given source. All contributions greater than 1\% were then added in quadrature. During this procedure, the average kinematic point in each bin was also recalculated. In all cases, the mean kinematics for a given bin did not vary significantly, and hence the associated systematic uncertainty was neglected. In the following, the sources of systematic uncertainty considered are presented and typical ranges are provided. A complete breakdown of the systematic uncertainties is available in the CLAS Physics Database~\cite{database}. The total systematic uncertainties are provided for all data points in \Cref{Appendix_Int_Res}, and the systematic uncertainties averaged over all photon-energy bins are given in \Cref{tab_syst}. 

\subsection{Bin-by-Bin Systematics}

\vskip 0.2cm
\noindent
\underline{Photon virtuality }: 
The impact of the cut applied to the reconstructed photon virtuality of the initial photon $\Qt2$ was tested by varying its value from 0.5~GeV$^2$ to 0.2~GeV$^2$ and 0.8~GeV$^2$. The efficiency correction factor $\omega_{c}$ was recalculated accordingly. The associated average systematic uncertainty was evaluated to be 6.53\%.

\vskip 0.2cm
\noindent
\underline{Missing mass squared}: 
The selection cut applied on the missing mass squared $\MM2$ was varied from 0.4 GeV$^2$ to 0.2 GeV$^2$ and 0.8 GeV$^2$. The corresponding average systematic uncertainty was determined to be 2.16\%.

\vskip 0.2cm
\noindent
\underline{Fit function}:
Three different combinations of signal and background functions were used to test their impact on the extraction of the number of J/$\psi$. The default fit used a Gaussian signal and a decaying exponential for the background. Variations included (i) a Crystal-Ball function~\cite{Skwarnicki:1986xj} for the signal with a decaying exponential background, and (ii) a Gaussian signal with a second-order polynomial for the background. For the Crystal-Ball signal, the parameters that described the left tail were fixed using the J$/\psi$ MC samples that included radiative effects to account for the radiative tail. The average systematic deviation associated with the choice of the fit function was found to be 13.36\%.

\vskip 0.2cm
\noindent
\underline{Lepton identification}:
The impact of the cut applied to the output of the BDT-based lepton-identification algorithm was evaluated by varying it around its nominal value. The associated average systematic uncertainty was found to be below 5\%.

\vskip 0.2cm
\noindent
\underline{Proton identification}:
The scattered proton was identified using the time-of-flight method. To assess the impact of the proton identification procedure, selection cuts at 2 and 3 standard deviations were applied to the distribution of the difference between the expected and measured proton time‑of‑flight, in both real data and MC. The corresponding average systematic uncertainty in the measured cross section was found to be below 5\%.

\vskip 0.2cm
\noindent
\underline{Spin Density Matrix Elements}:
The J$/\psi$ SDMEs have not been measured for exclusive near-threshold photoproduction, hence the cross section was extracted for three values of $\lambda_\theta$; 0.0, 0.5, and~1.0. The case $\lambda_\theta=1$, corresponding to $\rho^0_{00}=0$, was used by the GlueX Collaboration~\cite{PhysRevC.108.025201}. The case $\lambda_\theta=0$, corresponding to $\rho^0_{00}=1/3$, assumes an isotropic decay of the J$/\psi$ in its rest frame. Negative values of $\lambda_\theta$ ($\rho^0_{00}>1/3$) were not considered, as the photoproduction of longitudinally polarized vector mesons requires helicity-flip amplitudes that are dynamically suppressed (see e.g. Ref.~\cite{Mathieu:2018xyc}). The resulting uncertainty was close to 9\% and symmetric around $\lambda_\theta=0.5$, the value adopted in this analysis.

\vskip 0.2cm
\noindent
\underline{Radiative corrections}:
The radiative-correction factors were evaluated using the prescription of Ref.~\cite{Ehlotzky1968} and using the PHOTOS package~\cite{Barberio:1993qi}. The difference between the two approaches reached 10\% for low-energy bins, but was otherwise less than 5\%.

\subsection{Scale Systematics}

\vskip 0.2cm
\noindent
\underline{Virtual photon flux calculation}:
The use of a maximum value of the photon virtuality in the computation of the flux originated from the fact that JPsiGen does not include the scattered electron kinematics. Hence, the ratio of the total flux, where the virtuality is solely limited by the mass of the produced J$/\psi$, to the flux calculated with the virtuality limit should match the ratio between the acceptance-correction factor obtained using only JPsiGen or GRAPE, as the former does not include the photon virtuality dependence, while the latter does. This ratio was evaluated for all values of the $\Qt2$ cut. The largest discrepancy was found to be 4\%, which was assigned as the systematic uncertainty associated with the calculation of the virtual-photon flux.

\vskip 0.2cm
\noindent
\underline{Efficiency-correction factor}:
The systematic uncertainty associated with the efficiency-correction factor was set to 10.39\%, corresponding to the relative uncertainty of the efficiency-correction factor reported in \Cref{sec:eff_corr_factor}.  

\vskip 0.2cm
\noindent
\underline{Accumulated charge}:
A systematic uncertainty of 1.20\% was assigned to the determination of the accumulated charge. This estimation was obtained from the inclusive cross-section measurement using the same CLAS12 dataset in Ref.~\cite{CLAS:2025zup}.

\begin{table}[hbtp]
    \centering

\begin{ruledtabular}
\begin{tabular}{ l c } 
Average Bin-by-bin Systematics & [\%] \\
\noalign{\vskip 2pt}
\hline
\noalign{\vskip 2pt}
Photon virtuality & 6.53 \\
Missing mass squared & 2.16 \\
Fit function & 13.36 \\
Lepton identification & 4.54 \\
Proton identification & 4.88 \\
Spin Density Matrix Elements & 8.66 \\
Radiative corrections & 5.40 \\
\noalign{\vskip 2pt}
\hline\hline
\noalign{\vskip 2pt}
Scale Systematics & [\%] \\
\noalign{\vskip 2pt}
\hline
\noalign{\vskip 2pt}
Virtual photon flux calculation & 4.00 \\
Efficiency-correction factor & 10.39 \\
Accumulated charge & 1.20 \\
\end{tabular}
\end{ruledtabular}

\caption{Average systematic uncertainties over all photon energy bins. The dominant contributions come from the selection of the fit function, the treatment of SDMEs in the calculation of the acceptance factor, and the evaluation of the efficiency-correction factor.}
\label{tab_syst}
\end{table}

\section{Results and interpretation}
\label{sec:Results}

\begin{figure}[h!]
    \centering
    \includegraphics[width=\linewidth]{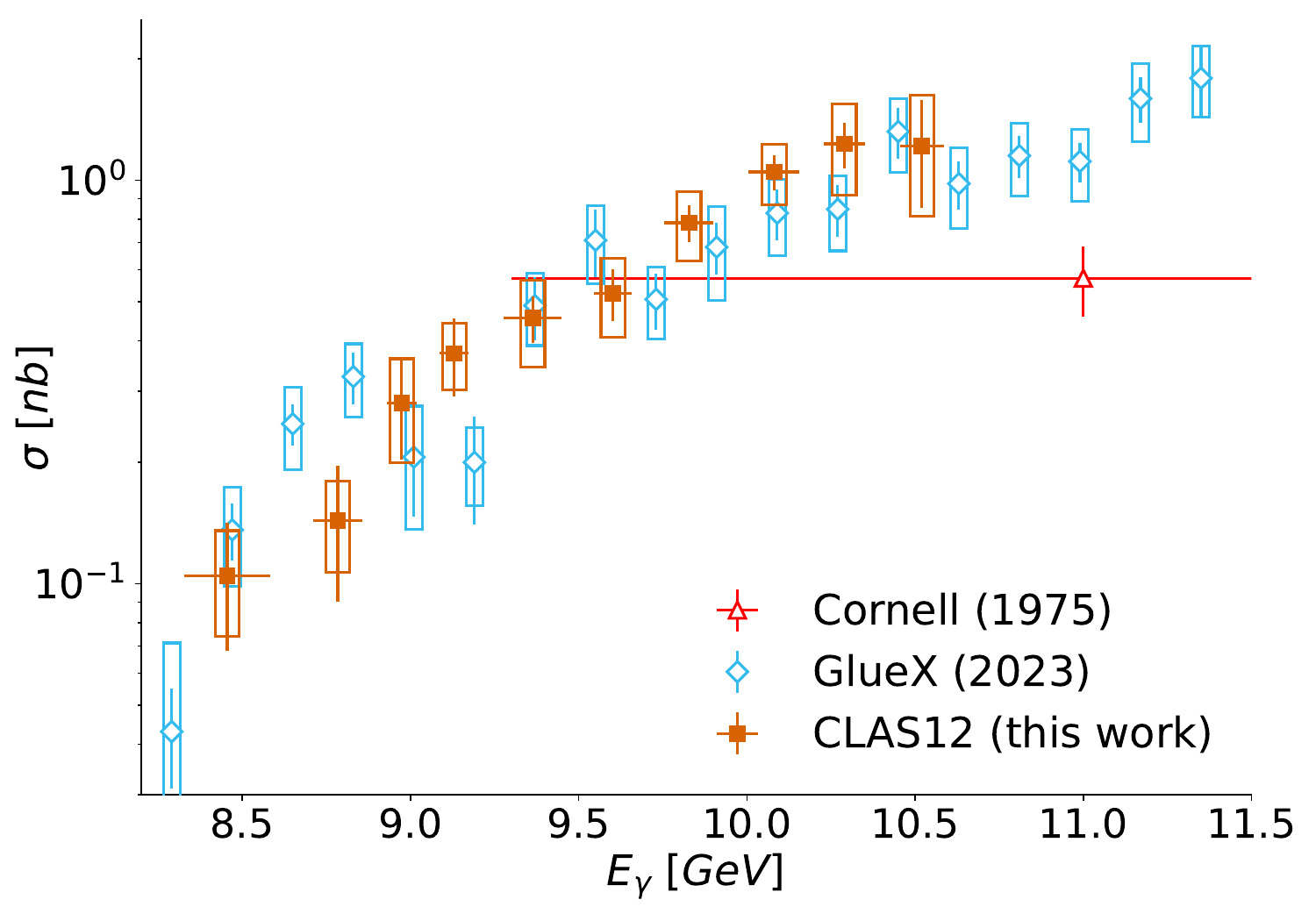}
    \caption{(Color online) Total cross sections for J$/\psi$ photoproduction measured with the CLAS12 experiment (orange squares). The results from the GlueX experiment~\cite{PhysRevC.108.025201} are shown as blue diamonds, and the single data point from Cornell~\cite{PhysRevLett.35.1616} is shown as a red triangle. Vertical error bars represent statistical uncertainties, while the boxes indicate systematic uncertainties. For the GlueX measurement, the 19.5\% normalization uncertainty reported in Ref.\cite{PhysRevC.108.025201} has been added in quadrature to the quoted systematic uncertainties.}
    \label{fig:Interp_Comparison_Int}
\end{figure}

The total cross sections for J$/\psi$ photoproduction measured with CLAS12 are shown in \Cref{fig:Interp_Comparison_Int}, compared with the Cornell data point~\cite{PhysRevLett.35.1616} and the GlueX measurement~\cite{PhysRevC.108.025201}. The di-muon results of the J$/\psi$-007 experiment~\cite{007:2026dow} were submitted to arXiv only after the CLAS12 analysis was concluded and without a publicly available dataset, and therefore could not be considered in the rest of this work. The scale of the measurements from the GlueX Collaboration and from this analysis agrees well, with comparable statistical and systematic uncertainties. In the vicinity of the $\Lambda_c\overline{D}$ and $\Lambda_c\overline{D}^*$ energy thresholds, around photon energies of 9~GeV, the GlueX Collaboration reported a potential dip of the cross section. This behavior has been discussed in models that include contributions from open-charm box-diagrams to the cross section~\cite{Du:2020bqj,PhysRevD.108.054018}. In contrast, the cross-section data measured in this work exhibit a smooth dependence on the photon energy over the entire range accessible by CLAS12, from threshold up to 10.6~GeV as shown in \Cref{fig:MCandData}. Additionally, the resolution of the reconstructed photon energy was evaluated using MC and found to be at least three times smaller than the bin width used in this analysis. Consequently, bin-migration effects were negligible and did not significantly distort the measured energy dependency.

\begin{figure}[h!]
    \centering
    \includegraphics[width=\linewidth]{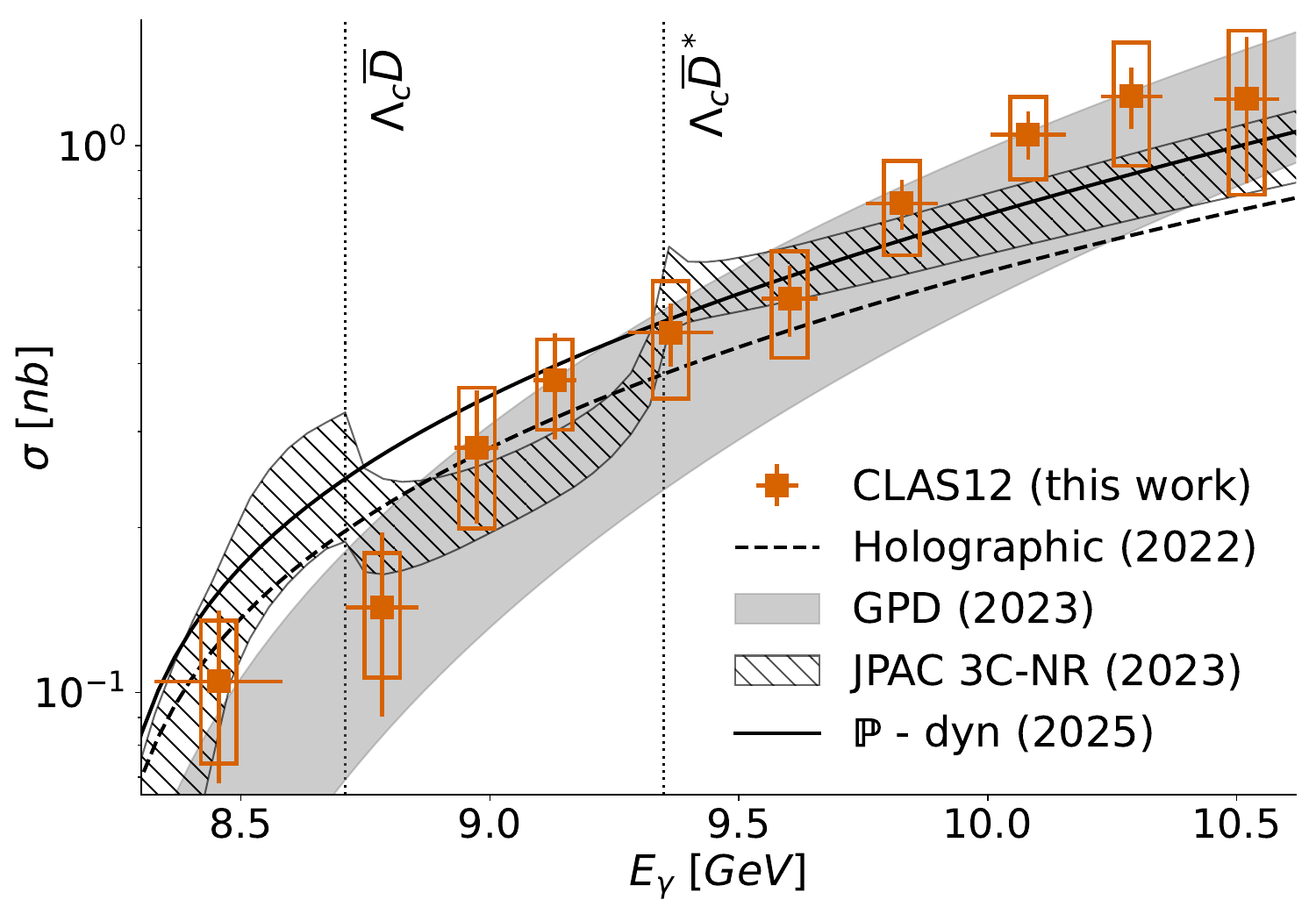}
    \caption{(Color online) Total cross sections for J$/\psi$ photoproduction measured with the CLAS12 experiment (orange squares). Vertical error bars represent statistical uncertainties, while the boxes indicate systematic uncertainties. The GPD model prediction~\cite{Guo:2023pqw} is shown as a gray band. The holographic-QCD model prediction~\cite{Mamo:2022eui} is displayed by the black dashed line. The model developed by the JPAC Collaboration in Ref.~\cite{PhysRevD.108.054018}, including open-charm contributions and denoted 3C-NR, is shown as the hashed band. The model based on Pomeron exchange proposed in Ref.~\cite{Tang:2025qqe}, denoted $\mathbb{P}-\rm dyn$, is displayed by the solid black line. The vertical dashed lines denote the mass thresholds of the $\Lambda_c\overline{D}$ and $\Lambda_c\overline{D}^*$ systems, respectively.}
    \label{fig:Interp_Int}
\end{figure}

In \Cref{fig:Interp_Int}, the CLAS12 results are compared with several model predictions: the GPD model from Ref.~\cite{Guo:2023pqw}, the holographic-QCD model from Ref.~\cite{Mamo:2022eui}, the model developed by the JPAC Collaboration in Ref.~\cite{PhysRevD.108.054018}, which includes open-charm contributions, and the Pomeron-based model from Ref.~\cite{Tang:2025qqe}. In all cases, the predictions are generally consistent with the data within two standard deviations.

\begin{figure}[hbtp]
    \centering
    \includegraphics[width=\linewidth]{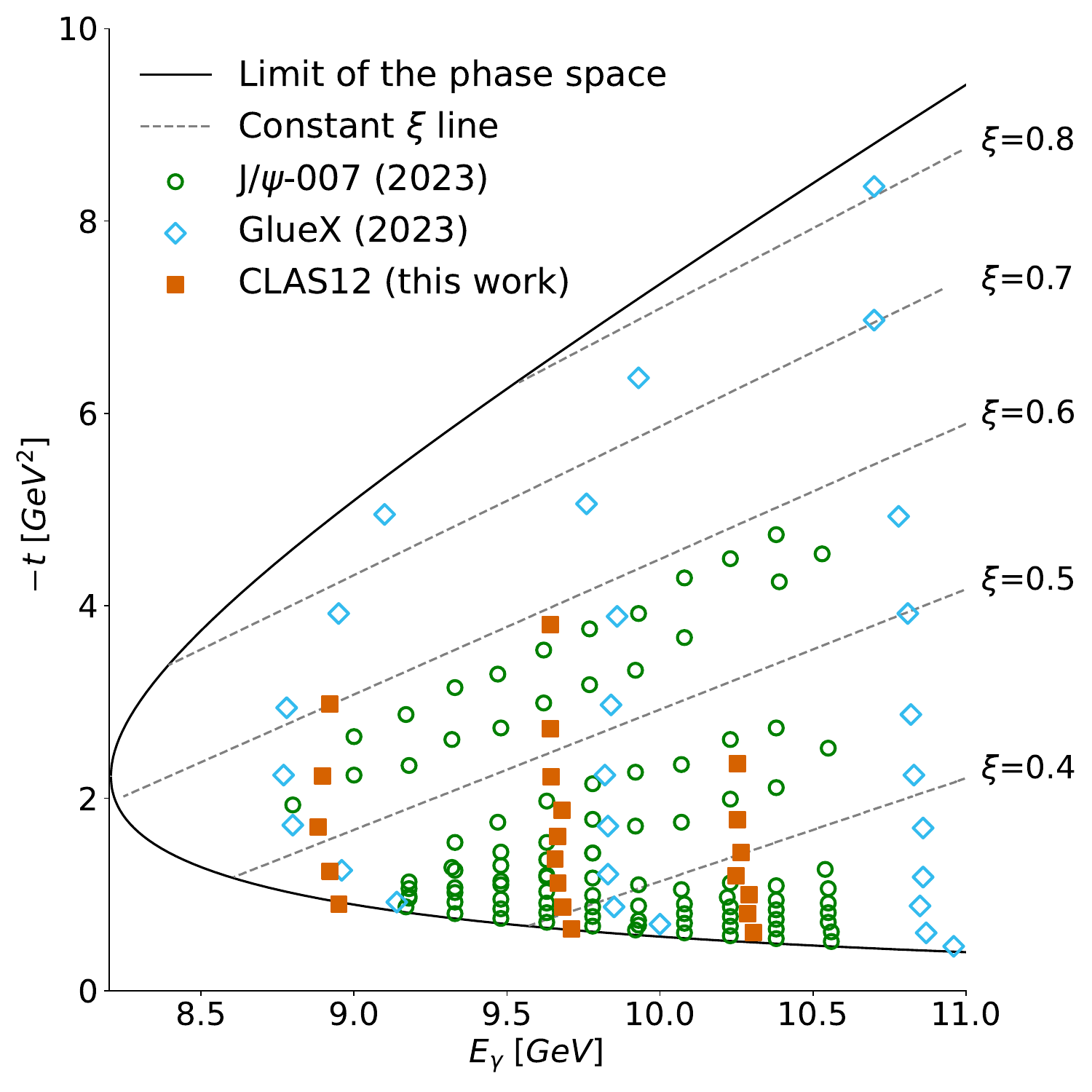}
    \caption{(Color online) Phase-space diagram for the near-threshold J$/\psi$ photoproduction, showing $-t$ as a function of the photon energy $E_\gamma$. The CLAS12 data points are displayed by orange squares. The GlueX~\cite{PhysRevC.108.025201} and J$/\psi$-007~\cite{Duran2023} points are shown in blue diamonds and green circles, respectively. The solid curve delimits the kinematically allowed reaction phase space. The dashed lines represent the locations of constant gluon momentum skewness in the GPD model, denoted by $\xi$.}
    \label{fig:Interp_Comparison_2D}
\end{figure}

The differential cross-section measurement of CLAS12 covers a phase space similar to that in the GlueX~\cite{PhysRevC.108.025201} and J$/\psi$-007~\cite{Duran2023} experiments. The mean values of $-t$ and $E_{\gamma}$ corresponding to each data point from the three experiments are presented in \Cref{fig:Interp_Comparison_2D}. Although the CLAS12 data do not extend to large $-t$ values due to the lack of lepton detection at very small polar angles, it provides new data points for gluon momentum skewness $\xi$ above 0.4 where the applicability of the GPD model is expected to be most reliable~\cite{Guo:2023pqw}. 

\begin{figure*}[hbtp]
    \centering
    \includegraphics[width=\linewidth]{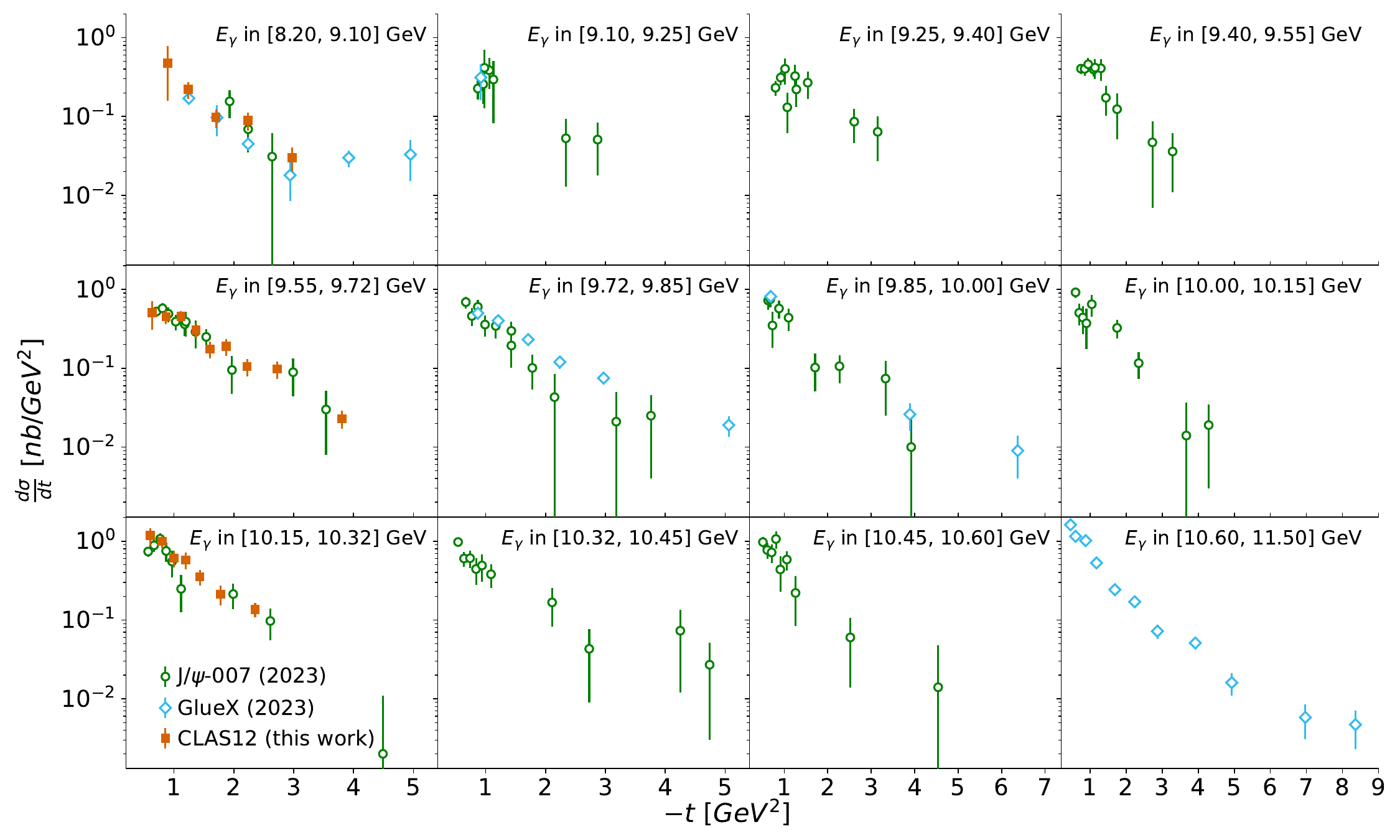}
    \caption{(Color online) Differential cross sections for J$/\psi$ photoproduction as a function of $-t$ for several photon energies. The CLAS12 data points are represented as orange squares, and compared with those from the GlueX experiment~\cite{PhysRevC.108.025201} (blue diamonds) and from the J$/\psi$-007 experiment~\cite{Du:2020bqj} (green circles).}
    \label{fig:All_Diff_Data}
\end{figure*}

The differential cross-section measurements obtained in this analysis are compared to those obtained by previous experiments in \Cref{fig:All_Diff_Data} for several bins of photon energy, and a good agreement is observed. 

To relate the $t$-dependence of the cross section to the gluon content of the proton, Refs.~\cite{KHARZEEV1999568,Frankfurt:2002ka,PhysRevD.104.054015} suggested a dipole-type parametrization of this dependence of the form  
\begin{equation}
\frac{d\sigma}{dt} = \left. \frac{d\sigma}{dt} \right| _0  \frac{1}{(1-t/m_S^2)^4}.
\end{equation}

\begin{table}[hbtp]
    \centering
    \begin{ruledtabular}
    \begin{tabular}{c c c c c}
    Energy bin & $\left. \frac{d\sigma}{dt}\right| _0$  & $m_S$ & $\sqrt{\langle r^2_m \rangle}$ & $\chi^2/\rm{ndf}$  \\
    \noalign{\vskip 2pt}
    $\rm [GeV]$  & [nb/GeV$^{2}$] & $\rm [GeV]$   & [fm] &   \\
    \noalign{\vskip 2pt}
    \hline
    \noalign{\vskip 2pt}
    $[8.20, 9.28]$ & 2.04$\pm$1.47 & 1.28$\pm$0.22 & 0.53$\pm$0.09 & 0.64 \\ 
    $[9.28, 10.00]$ & 2.24$\pm$0.64 & 1.39$\pm$0.11 & 0.49$\pm$0.04 & 1.10 \\ 
    $[10.00, 10.60]$ & 4.25$\pm$0.67 & 1.30$\pm$0.06 & 0.53$\pm$0.02 & 0.21 \\
    \end{tabular}
    \caption{Results of the dipole fit of the CLAS12 differential cross section for the three photon-energy bins. Reported uncertainties are statistical only.}
    \label{tab:dipole_fit}
    \end{ruledtabular}
\end{table}

\begin{figure}[hbtp]
    \centering
    \includegraphics[width=\linewidth]{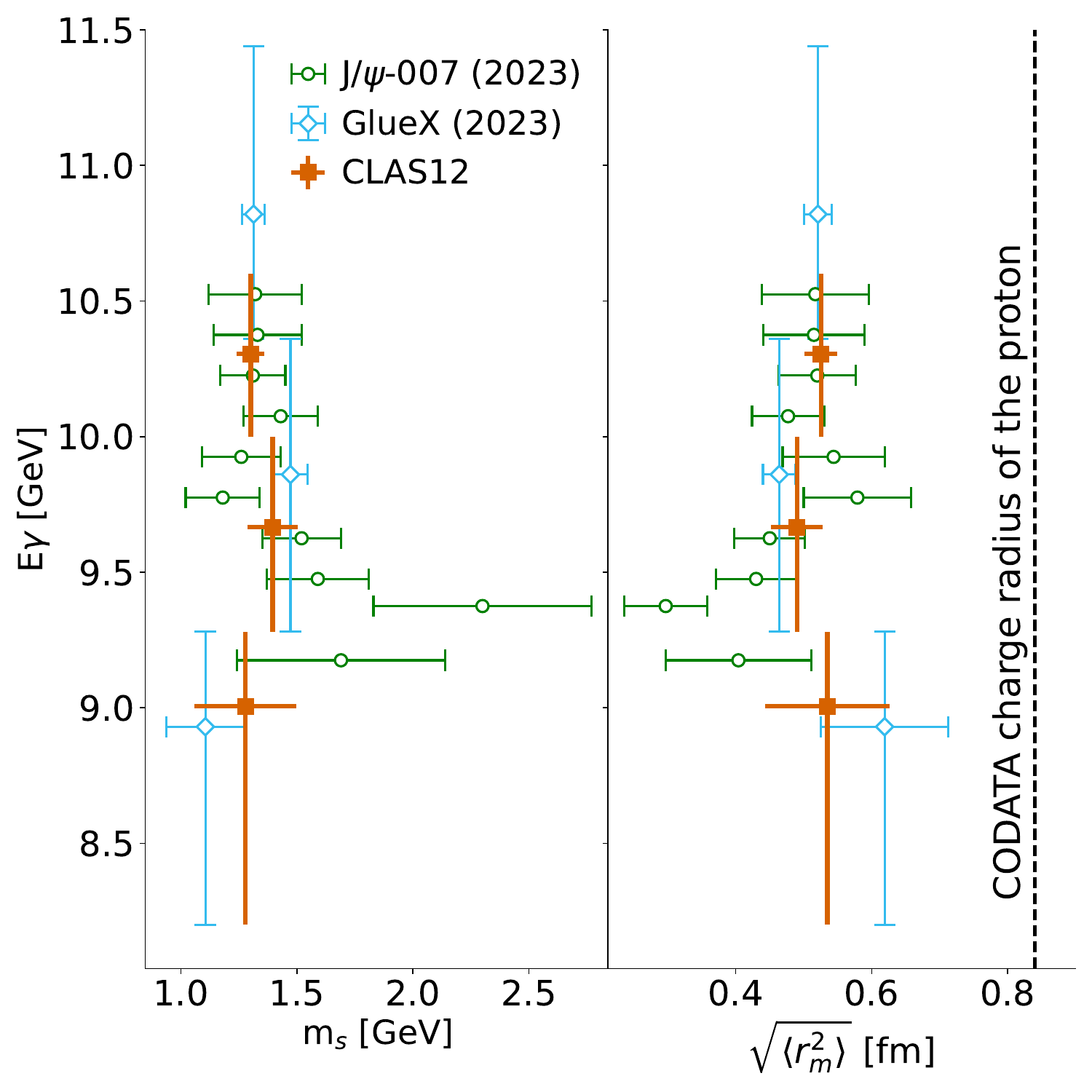}
    \caption{(Color online) Slope parameter of the dipole model $m_S$ (left) and the corresponding mass radius of the proton $\sqrt{\langle r^2_m \rangle}$ (right), as a function of $E_{\gamma}$. The CLAS12 results are shown as orange squares, previous results from the GlueX~\cite{PhysRevC.108.025201} and J$/\psi$-007~\cite{Duran2023} experiments are shown in blue diamonds and green circles, respectively. For comparison, the CODATA-2018~\cite{Tiesinga:2021myr} proton charge radius is indicated by the vertical dashed line.}
    \label{fig:Interp_Mass_rad}
\end{figure}

\Cref{tab:dipole_fit} shows the parameters of the dipole fits to the CLAS12 data. According to the prescription in Ref.~\cite{PhysRevD.104.054015}, the parameter $m_S$ is related to the mass radius of the proton as
\begin{equation}
\sqrt{\langle r^2_m \rangle}=\frac{\sqrt{12}}{m_S}.
\end{equation}
Figure~\ref{fig:Interp_Mass_rad} shows the mass radii obtained from the differential cross sections measured with CLAS12 data, compared to the values extracted by the GlueX and J$/\psi$-007 experiments. The mass radii obtained from the CLAS12 data are consistent with previously reported results. Close to the production threshold, where the dipole model is expected to be the most reliable, only GlueX measurements were available prior to this work. The present results confirm the GlueX extraction and indicate a proton mass radius of the order of $\sim 0.5$ fm, smaller than the proton charge radius.

Using the GPD~\cite{Guo:2023pqw} and holographic-QCD~\cite{Mamo:2022eui} models, the GFFs $A_g(t)$ and $C_g(t)$ were extracted from the CLAS12 data. Following the approaches in Ref.~\cite{Duran2023}, a tripole ansatz was used, giving

\begin{equation}
    A_g(t)= \frac{A_g(0)}{\left( 1 - \frac{t}{m^2_A}\right)^3},~
    C_g(t)= \frac{C_g(0)}{\left( 1 - \frac{t}{m^2_C}\right)^3},
\end{equation}
where $m_A$, $C_g(0)$, and $m_C$ are free parameters. The value of $A_g(0)$, which is related to the fraction of proton momentum carried by the gluons, was fixed to $0.414 \pm 0.008$ based on the CT18 global fit~\cite{Hou:2019efy}. The GFF $B_g(t)$ was ignored, following lattice-QCD indications~\cite{Pefkou:2021fni,Hackett:2023rif}. The GFF $\bar C_g(t)$ was also not considered, as it remains poorly constrained by lattice QCD, although its size could be large~\cite{Tanaka:2018nae, Tanaka:2022wzy} since $\bar C(0)$ is related to the trace anomaly of the QCD energy-momentum tensor. For the GPD model, only data points with sufficiently large $\xi$, the imbalance of the fraction of the longitudinal momentum of the proton carried by the probed gluons, were included in the fit ($\xi>0.4$), as suggested in Ref.~\cite{Guo:2023pqw}. In total, four fits were performed, two for each model, using the CLAS12 data alone or combining all existing data. When fitting only the CLAS12 data, the $m_C$ parameter was allowed to vary, and a Gaussian constraint was added to the fit residual. In the case of the GPD model, the prior value of $m_C$ was set to $0.91\pm0.10~{\rm GeV}$ following the results reported in Ref.~\cite{Guo:2023pqw}. For the holographic-QCD model, $m_C=1.12\pm0.21~{\rm GeV}$~was used, as reported in Ref.~\cite{Duran2023}. The fitted parameters obtained for the different models and datasets are summarized in \Cref{tab:results_fit_GFFs}, including the posterior value of $m_C$. The correlations between the fitted parameters, which were accounted for in the calculation of the uncertainties of the GFFs, are reported in \Cref{tab:results_fit_corr} in \Cref{Appendix_Corr}.

\begin{table}[hbtp]
    \centering
    \begin{ruledtabular}
    \begin{tabular}{c c c}
    \noalign{\vskip 1pt}
  \multicolumn{3}{c}{GPD model} \\
  \noalign{\vskip 2pt}
  \hline
  \noalign{\vskip 2pt}
Dataset 
& CLAS12 
& Combined
\\

$\chi^2/\rm{ndf}$
& 1.29
& 1.37
\\

$A_g(0)$
& [$0.414 \pm 0.008$]
& [$0.414 \pm 0.008$]
\\

$m_A$ [GeV]
& $2.12 \pm 0.08$
& $2.08 \pm 0.06$
\\

$C_g(0)$
& -$1.36 \pm 0.52$
& -$1.32 \pm 0.52$
\\

$m_C$ [GeV]
& ($0.91 \pm 0.10$) $0.89 \pm 0.11$ 
& $0.88 \pm 0.12$
\\

\noalign{\vskip 2pt}

\hline\hline
\noalign{\vskip 2pt}
 \multicolumn{3}{c}{Holographic model} \\
 \noalign{\vskip 2pt}
 \hline
 \noalign{\vskip 2pt}
Dataset 
& CLAS12 
& Combined \\

$\chi^2/\rm{ndf}$
& 0.62
& 1.05 \\

$A_g(0)$
& [$0.414 \pm 0.008$]
& [$0.414 \pm 0.008$] \\

$m_A$ [GeV]
& $1.64 \pm 0.05$
& $1.70 \pm 0.032$ \\

$C_g(0)$
& -$0.38 \pm 0.11$
& -$0.38 \pm 0.03$ \\

$m_C$ [GeV]
& ($1.12 \pm 0.21$) $1.12 \pm 0.15$ 
& $1.38 \pm 0.09$ \\
\end{tabular}

    \end{ruledtabular}
    \caption{Summary of the fitted GFF parameters obtained using only the CLAS12 data, and combining the GlueX, J$/\psi$-007, and CLAS12 datasets, for the GPD~\cite{Guo:2023pqw} and the holographic-QCD~\cite{Mamo:2022eui} models. For all fits, the $A_g(0)$ parameter was fixed to the value within brackets. When fitting to the CLAS12 data alone, the prior values of $m_C$ are reported in parentheses. Statistical uncertainties of the fitted parameters are reported.}
    \label{tab:results_fit_GFFs}
\end{table}

\begin{figure}[htp]
  \centering
   \includegraphics[width=\linewidth]{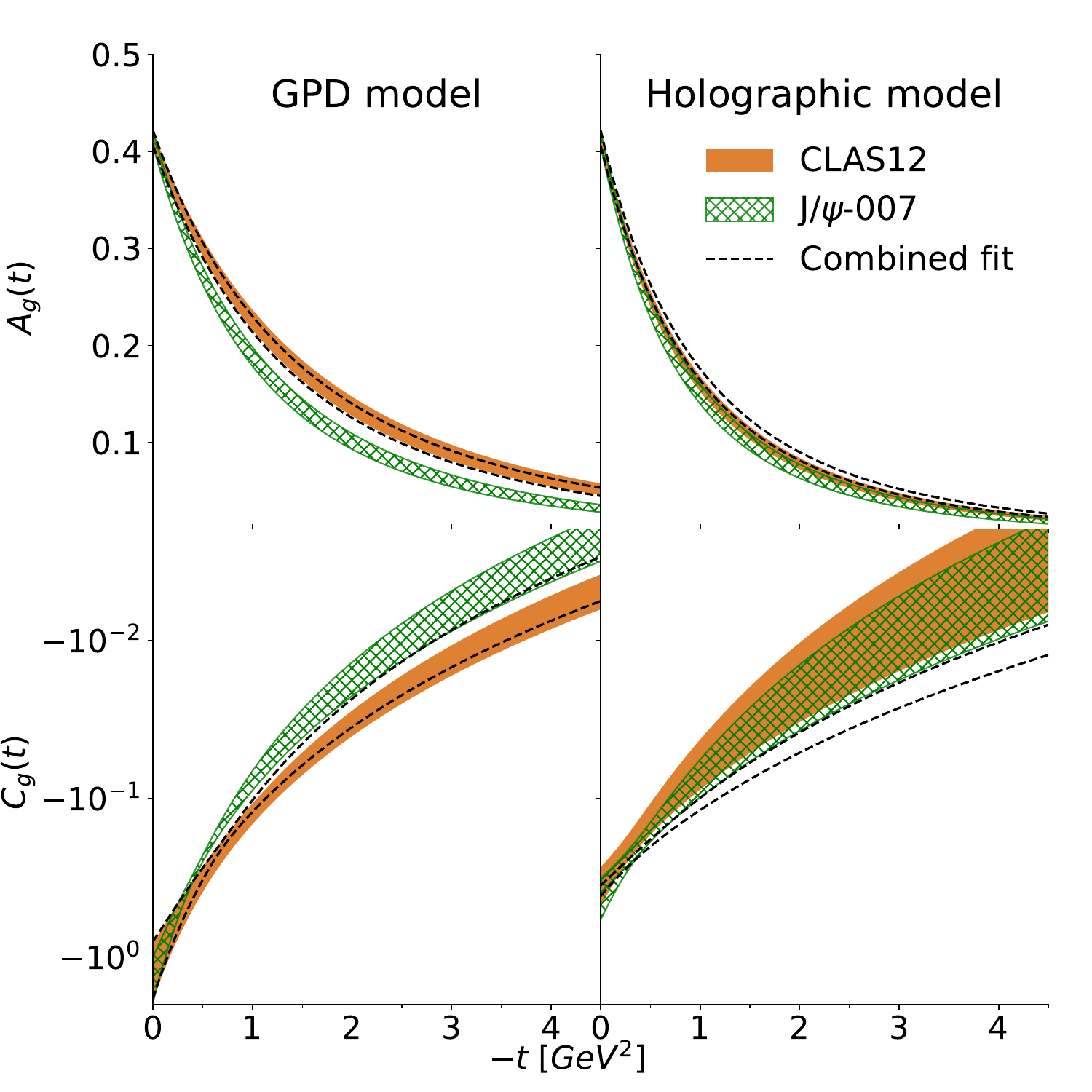}
\caption{(Color online) $A_g(t)$ and $C_g(t)$ Gravitational Form Factors for the GPD and holographic models. The fit using only CLAS12 data is shown in orange, the results obtained by the \mbox{J$/\psi$-007} Collaboration in Refs.~\cite{Duran2023,Meziani:2024cke} are shown in hashed green, and the combined fit including the GlueX, J$/\psi$-007, and CLAS12 data is shown with two dashed lines representing the minimum and maximum of the statistical uncertainties. Error bars include the correlations between the fitted parameters reported in \Cref{tab:results_fit_corr} in \Cref{Appendix_Corr}.}
\label{fig:Interp_GFFs}
\end{figure}

The GFFs $A_g(t)$ and $C_g(t)$ extracted in this work are shown in \Cref{fig:Interp_GFFs}. For comparison, the GFFs extracted from the J$/\psi$-007 data are superimposed, using the holographic-QCD extraction from Ref.~\cite{Duran2023} and the revised GPD analysis reported in Ref.~\cite{Meziani:2024cke}. For both theoretical models, the CLAS12-only and combined extractions of $A_g$ tend to favor larger values of $m_A$ compared to those reported by the J$/\psi$-007 experiment. Although $m_C$ was constrained in the fits that used only CLAS12 data, the obtained values of $C_g(0)$ are consistent with those extracted by the J$/\psi$-007 Collaboration~\cite{Meziani:2024cke}. The GPD model favors a higher absolute value of $C_g(0)$, which translates into a higher pressure at the core of the proton for a given value of $m_C$. In the holographic-QCD model case, the combined extraction of the $C_g(t)$ form factor exhibits a flatter $t$-dependence, which is driven mainly by the large $|t|$ data points of the GlueX experiment.

The transverse and shear pressure distributions produced by the gluons in the proton can be inferred from the Fourier transform of the $C_g(t)$ form factor following the prescription in Refs.~\cite{Polyakov:2018zvc, Lorce:2018egm}. In the following, we instead present the mass radius $\langle r_{m}^2\rangle_g$ and the scalar radius $\langle r_{s}^2\rangle_g $ of the proton defined in Ref.~\cite{Ji:2021mtz} as

\begin{align}
\label{eqn:m_radius}
\langle r_{m}^2\rangle_g
&= \left. 6\,\frac{1}{A_g(0)} \frac{dA_g(t)}{dt} \right|_{t=0}
   - 6\,\frac{1}{A_g(0)} \frac{C_g(0)}{m_p^2} \notag \\
&= \frac{18}{m_A^2}
   - 6\,\frac{1}{A_g(0)} \frac{C_g(0)}{m_p^2} ,
\end{align}

\begin{align}
\label{eqn:s_radius}
\langle r_{s}^2\rangle_g
&= \left. 6\,\frac{1}{A_g(0)} \frac{dA_g(t)}{dt} \right|_{t=0}
   - 18\,\frac{1}{A_g(0)} \frac{C_g(0)}{m_p^2} \notag \\
&= \frac{18}{m_A^2}
   - 18\,\frac{1}{A_g(0)} \frac{C_g(0)}{m_p^2} ,
\end{align}
where $m_p$ is the mass of the proton, and the second equalities in each equation correspond to the case where a tripole $t$-dependence was used. 

\begin{figure}[h]
    \centering
    \includegraphics[width=\linewidth, page = 1]{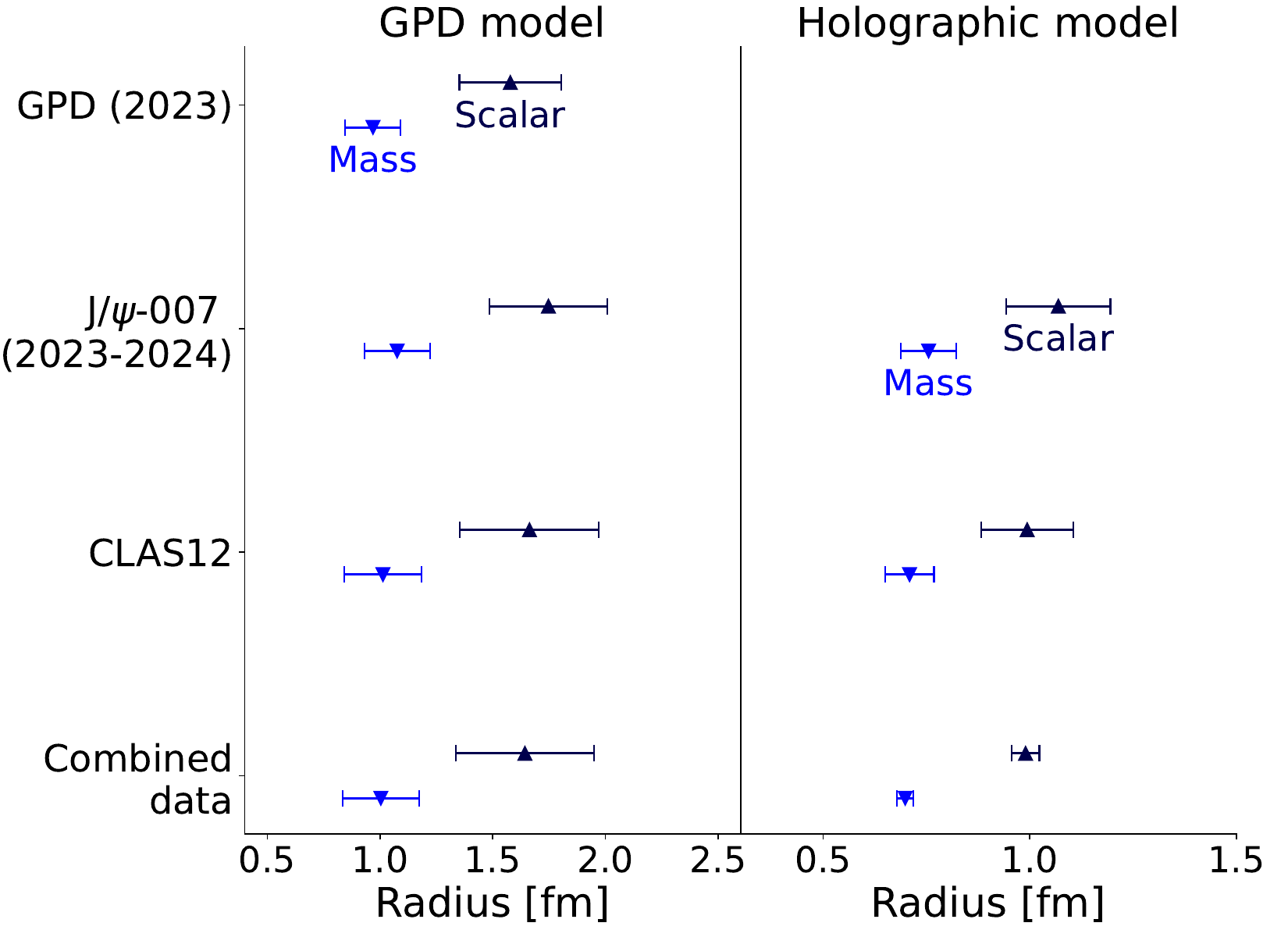}
    \caption{(Color online) Extracted mass and scalar radii from the CLAS12 data only, and from the combined GlueX, J$/\psi$-007, and CLAS12 datasets. The radii extracted with the GPD model in Ref.~\cite{Guo:2023pqw} are shown in the first line. The radii extracted from the J$/\psi$-007 dataset in Ref.~\cite{Meziani:2024cke} for the GPD model and in Ref.~\cite{Duran2023} for the holographic-QCD model are shown in the second line.}
    \label{fig:Interp_Radii}
\end{figure}

Figure~\ref{fig:Interp_Radii} shows the proton mass and scalar radii extracted in this analysis, together with previous extractions from the J$/\psi$-007 data \cite{Duran2023,Meziani:2024cke} and from Ref.~\cite{Guo:2023pqw}. \Cref{tab:radii} summarizes the values of the radii extracted in this work. For both cross-section models, the radii obtained from the CLAS12 data alone are compatible with previous results, but favor slightly smaller radii due to the larger fitted values of $m_A$. Including all available data into the fit resulted in values of radii compatible with all previous extractions, with reduced uncertainties for the case of the holographic model.

\begin{table}[hbtp]
    \centering
    \begin{ruledtabular}
    \begin{tabular}{c c c c}
    Model & Dataset & $\sqrt{\langle r^2_m\rangle}_g$ [fm] & $\sqrt{\langle r^2_s\rangle}_g$ [fm] \\
    \noalign{\vskip 2pt}
    \hline
    \noalign{\vskip 2pt}
    \multirow{2}{*}{GPD} & CLAS12 & $1.01 \pm 0.17$ & $1.66 \pm 0.31$ \\
                          & Combined data & $1.00 \pm 0.17$ & $1.64 \pm 0.31$ \\
    \noalign{\vskip 2pt}
    \hline
    \noalign{\vskip 2pt}
    \multirow{2}{*}{Holographic} & CLAS12 & $0.71 \pm 0.06$ & $0.99 \pm 0.11$ \\
                                  & Combined data & $0.70 \pm 0.02$ & $0.99 \pm 0.03$ \\
    \end{tabular}
    \end{ruledtabular}
    \caption{Extracted mass and scalar radii from the CLAS12 data only, and using the combined GlueX, J$/\psi$-007, and CLAS12 datasets. Statistical uncertainties are reported.}
    \label{tab:radii}
\end{table}

\section{Conclusions}
\label{sec:Conclusions}

In this article, data collected with the CLAS12 experiment at the Thomas Jefferson National Accelerator Facility have been analyzed to provide new and independent measurements of the total and differential cross sections for J$/\psi$ photoproduction near threshold. The total cross section was found to be consistent with previous results from the GlueX Collaboration. However, in the energy region in the vicinity of the open-charm thresholds, the CLAS12 data exhibit a smoother energy dependence, which may indicate a small contribution from open-charm channels to the total cross section. New differential cross-section data have also been presented. These results are in good agreement with existing measurements from the GlueX and J$/\psi$-007 experiments and provide additional constraints close to the production threshold. An interpretation of these data in terms of the gluon content of the proton has been provided. In particular, the gluon Gravitational Form Factors $A_g(t)$ and $C_g(t)$ of the proton have been extracted using both GPD and holographic-QCD approaches. This analysis provides important new data points, but the statistical precision and $t$-coverage remain comparable to those of previous measurements. Future high-luminosity measurements at JLab with the SoLID experiment~\cite{E12-12-006, JeffersonLabSoLID:2022iod} and the recently approved $\rm \mu CLAS12$ setup~\cite{Alvarado:2026ggy} will provide significantly larger samples of J$/\psi$ photoproduction events and substantially improve the precision of near-threshold J$/\psi$ production studies. Further ahead, the Electron-Ion Collider at Brookhaven National Laboratory will allow measurements of exclusive photoproduction of both J$/\psi$ and $\Upsilon$~\cite{Gryniuk:2020mlh}, providing a unique understanding of the gluon structure of the proton.

\section*{Acknowledgments}
We thank Z-E.~Meziani and S.~Prasad for fruitful discussions on the GFF fits; L.~Pentchev for providing the extrapolated Cornell data point; G.~Monta\~na, V.~Mathieu, and D.~Winney for discussions about the SDME formalism and for providing the JPAC model predictions; and C.~Roberts for providing the prediction for the total cross section based on the Pomeron exchange model. The authors acknowledge the outstanding efforts of the staff of the Accelerator and Physics Divisions at Jefferson Lab in making this experiment possible. 

This work was supported in part by the U.S. Department of Energy, the National Science Foundation (NSF), and the Italian Istituto Nazionale di Fisica Nucleare (INFN), the French Centre National de la Recherche Scientifique (CNRS), the French Commissariat à l’Energie Atomique (CEA), the UK Science and Technology Facilities Council (STFC), the National Research Foundation of Korea (NRF), the Helmholtz-Forschungsakademie
Hessen für FAIR (HFHF), the Chilean Comisión Nacional de Investigación Científica y Tecnológica (CONICYT), the Scottish Universities Physics Alliance (SUPA), the Skobeltsyn Nuclear Physics Institute and Physics Department at the Lomonosov Moscow State University. This work was
also supported by the Marie Sklodowska-Curie Grant Agreement No.~101003460. This material is based on work supported by the U.S. Department of Energy, Office of Science, Office of Nuclear Physics under Contract No.~DE-AC05-06OR23177.

\appendix

\section{Numerical Results}
\label{Appendix_Int_Res}
The numerical values of the total cross section as a function of the incoming photon energy are given in \Cref{tab:Tab_res_int}.

\begin{table}[h!]
\centering
\begin{ruledtabular}
\begin{tabular}{c c c}
Energy bin [GeV] & $\langle E_\gamma \rangle$ [GeV] & $\sigma$ [nb] \\
\noalign{\vskip 2pt}
\hline
\noalign{\vskip 2pt}
$[8.20,\,8.65]$ & $8.45 \pm 0.13$  & $0.107 \pm 0.037 \pm 0.031$ \\
$[8.65,\,8.90]$ & $8.78 \pm 0.07$  & $0.146 \pm 0.051 \pm 0.037$ \\
$[8.90,\,9.05]$ & $8.97 \pm 0.04$  & $0.280 \pm 0.076 \pm 0.080$ \\
$[9.05,\,9.20]$ & $9.13 \pm 0.04$  & $0.373 \pm 0.081 \pm 0.070$ \\
$[9.20,\,9.50]$ & $9.36 \pm 0.09$  & $0.456 \pm 0.060 \pm 0.110$ \\
$[9.50,\,9.70]$ & $9.60 \pm 0.06$  & $0.526 \pm 0.078 \pm 0.116$ \\
$[9.70,\,10.00]$ & $9.82 \pm 0.07$ & $0.783 \pm 0.083 \pm 0.153$ \\
$[10.00,\,10.20]$ & $10.08 \pm 0.07$ & $1.050 \pm 0.106 \pm 0.181$ \\
$[10.20,\,10.40]$ & $10.29 \pm 0.06$ & $1.233 \pm 0.159 \pm 0.312$ \\
$[10.40,\,10.60]$ & $10.52 \pm 0.06$ & $1.220 \pm 0.363 \pm 0.405$ \\
\end{tabular}
\end{ruledtabular}
\caption{Values of the total cross sections for J$/\psi$ photoproduction measured with the CLAS12 experiment in bins of photon energy. The energy range and average energy with its standard deviation are given for each bin. The first and second uncertainties correspond to statistical and systematic uncertainties, respectively.}
\label{tab:Tab_res_int}
\end{table}

The numerical values of the differential cross section as a function of the Mandelstam variable $-t$ are given in \Cref{tab:Tab_res_diff}.

\begin{table}[ht]
    \centering

    \begin{ruledtabular}
\begin{tabular}{c c c c}
$-t$ bin & $\langle -t \rangle$ & $\langle E_\gamma \rangle$  & ${d\sigma}/{dt}$  \\
\noalign{\vskip 2pt}
[GeV$^2$] & [GeV$^2$] & [GeV] & [nb/GeV$^{2}$] \\
\noalign{\vskip 2pt}
\hline
\noalign{\vskip 2pt}
$[0.77,\,1.00]$ & $0.90 \pm 0.06$ & $8.95 \pm 0.26$ & $0.473 \pm 0.316 \pm 0.225$ \\
$[1.00,\,1.50]$ & $1.24 \pm 0.14$ & $8.92 \pm 0.26$ & $0.221 \pm 0.054 \pm 0.044$ \\
$[1.50,\,2.00]$ & $1.70 \pm 0.14$ & $8.88 \pm 0.27$ & $0.097 \pm 0.026 \pm 0.023$ \\
$[2.00,\,2.50]$ & $2.23 \pm 0.15$ & $8.90 \pm 0.28$ & $0.089 \pm 0.023 \pm 0.019$ \\
$[2.50,\,4.50]$ & $2.98 \pm 0.43$ & $8.92 \pm 0.25$ & $0.030 \pm 0.010 \pm 0.008$ \\
\noalign{\vskip 2pt}
\hline
\noalign{\vskip 2pt}
$[0.50,\,0.75]$ & $0.64 \pm 0.07$ & $9.71 \pm 0.18$ & $0.506 \pm 0.201 \pm 0.146$ \\
$[0.75,\,1.00]$ & $0.87 \pm 0.07$ & $9.68 \pm 0.21$ & $0.451 \pm 0.088 \pm 0.100$ \\
$[1.00,\,1.25]$ & $1.12 \pm 0.07$ & $9.67 \pm 0.20$ & $0.457 \pm 0.070 \pm 0.073$ \\
$[1.25,\,1.50]$ & $1.37 \pm 0.07$ & $9.66 \pm 0.20$ & $0.305 \pm 0.057 \pm 0.080$ \\
$[1.50,\,1.75]$ & $1.60 \pm 0.07$ & $9.67 \pm 0.21$ & $0.175 \pm 0.043 \pm 0.029$ \\
$[1.75,\,2.00]$ & $1.87 \pm 0.08$ & $9.68 \pm 0.19$ & $0.189 \pm 0.045 \pm 0.055$ \\
$[2.00,\,2.50]$ & $2.22 \pm 0.15$ & $9.65 \pm 0.21$ & $0.105 \pm 0.026 \pm 0.022$ \\
$[2.50,\,3.00]$ & $2.72 \pm 0.14$ & $9.64 \pm 0.22$ & $0.098 \pm 0.025 \pm 0.024$ \\
$[3.00,\,6.00]$ & $3.81 \pm 0.68$ & $9.64 \pm 0.19$ & $0.023 \pm 0.006 \pm 0.004$ \\
\noalign{\vskip 2pt}
\hline
\noalign{\vskip 2pt}
$[0.50,\,0.70]$ & $0.60 \pm 0.06$ & $10.30 \pm 0.19$ & $1.176 \pm 0.291 \pm 0.356$ \\
$[0.70,\,0.90]$ & $0.80 \pm 0.06$ & $10.29 \pm 0.17$ & $0.991 \pm 0.179 \pm 0.300$ \\
$[0.90,\,1.10]$ & $1.00 \pm 0.06$ & $10.29 \pm 0.19$ & $0.609 \pm 0.143 \pm 0.179$ \\
$[1.10,\,1.30]$ & $1.19 \pm 0.06$ & $10.25 \pm 0.17$ & $0.575 \pm 0.137 \pm 0.156$ \\
$[1.30,\,1.60]$ & $1.43 \pm 0.08$ & $10.27 \pm 0.18$ & $0.354 \pm 0.078 \pm 0.135$ \\
$[1.60,\,2.00]$ & $1.78 \pm 0.10$ & $10.25 \pm 0.18$ & $0.213 \pm 0.060 \pm 0.037$ \\
$[2.00,\,3.00]$ & $2.36 \pm 0.26$ & $10.25 \pm 0.17$ & $0.136 \pm 0.027 \pm 0.034$ \\
\end{tabular}
\end{ruledtabular}

\caption{Values of the differential cross sections for J$/\psi$ photoproduction measured with the CLAS12 experiment in bins of $-t$. The $-t$ range, average $-t$ value with its standard deviation, and average photon energy with its standard deviation are given for each bin. Each sub-table corresponds to a photon-energy bin: [8.2, 9.28] GeV, [9.28, 10.00] GeV, and [10.0, 10.6] GeV. The first and second uncertainties correspond to the statistical and systematic uncertainties, respectively.}
    \label{tab:Tab_res_diff}
\end{table}

\section{Correlations of the Fitted GFF Parameters}
\label{Appendix_Corr}

\Cref{tab:results_fit_corr} summarizes the correlation values for the fitted parameters of the GFFs, entering in the computation of the uncertainties of the extracted GFFs presented in \Cref{sec:Results}.

\begin{table}[ht]
    \centering
    \begin{ruledtabular}
    \begin{tabular}{c c c c}
    Model & Dataset & $\rho(m_A,C(0))$ & $\rho(m_C,C(0))$ \\
    \noalign{\vskip 2pt}
    \hline
    \noalign{\vskip 2pt}
    \multirow{2}{*}{GPD} & CLAS12 & -0.88 & 0.92 \\
                         & All data & 0.82 & 0.99 \\
    \noalign{\vskip 2pt}
    \hline
    \noalign{\vskip 2pt}
    \multirow{2}{*}{Holographic} & CLAS12 & -0.91 & 0.73 \\
                                 & All data & 0.39 & 0.75 \\
    \end{tabular}
    \caption{Correlations of the fitted GFF parameters entering in the computation of the uncertainties of the extracted GFFs presented in \Cref{sec:Results}. }
    \label{tab:results_fit_corr}
    \end{ruledtabular}
\end{table}


%

\end{document}